\documentclass[12pt,preprint]{aastex}

\slugcomment{To appear in ApJ.}
\shorttitle{MHD Outflow Dynamics}
\shortauthors{Hawley, Krolik}
\usepackage{psfig}

\received{2005 July 25}
\begin{document}

\title{Magnetically Driven Jets in the Kerr Metric }


\author{John F. Hawley}
\affil{Astronomy Department\\
University of Virginia\\ 
P.O. Box 3818, University Station\\
Charlottesville, VA 22903-0818}

\and
\author{Julian H. Krolik}
\affil{Department of Physics and Astronomy\\
Johns Hopkins University\\
Baltimore, MD 21218}

\email{jh8h@virginia.edu; jhk@pha.jhu.edu}

\begin{abstract}

We compute a series of three-dimensional general relativistic
magnetohydrodynamic simulations of accretion flows in the Kerr metric
to investigate the properties of the unbound outflows that result.
The overall strength of these outflows increases sharply with increasing
black hole rotation rate, but a number of generic features are found
in all cases.  The mass in the outflow is concentrated in a hollow cone
whose opening angle is largely determined by the effective potential for
matter orbiting with angular momentum comparable to that of the innermost
stable circular orbit.  The dominant force accelerating the matter outward
comes from the pressure of the accretion disk's corona.  The principal
element that shapes the outflow is therefore the centrifugal barrier
preventing accreting matter from coming close to the rotation axis.
Inside the centrifugal barrier, the cone contains very little matter and
is dominated by electromagnetic fields.  The magnetic fieldlines inside
the cone rotate at a rate tied closely to the rotation of the black hole,
even when the black hole spins in a sense opposite to the rotation of the
accretion flow.  These fields carry an outward-going Poynting flux whose
immediate energy source is the rotating spacetime of the Kerr black hole.
When the spin parameter $a/M$ of the black hole exceeds $\simeq 0.9$,
the energy carried to infinity by these outflows can be comparable to
the nominal radiative efficiency predicted in the Novikov-Thorne model.
Similarly, the expelled angular momentum can be comparable to that
accreted by the black hole.  Both the inner electromagnetic part and the
outer matter part can contribute in significant fashion to the energy
and angular momentum of the outflow.

\end{abstract}


\section{Introduction}

     Whether from supermassive objects in galactic nuclei or stellar mass
objects in Galactic binaries, outflows---often relativistic---are
frequently seen from black holes.  However, there is little that is
known with confidence about their nature and dynamics.  The essential
ingredients are, of course, well-known:  magnetic field, accretion, and
rotation.  But basic questions such as the forces that drive the jets,
the mechanisms that regulate their content, and the constraints that
collimate them remain open.  Magneto-centrifugal effects have received
the greatest attention (Blandford \& Znajek 1977; Blandford \& Payne 1981;
Shibata \& Uchida 1985; Punsly \& Coroniti 1990), but there are many
possible variations on this idea.  In most such studies, magnetic field
boundary conditions are guessed and their consequences derived without
much consideration of how the posited field structures would arise in
the context of accretion flows.

We approach these questions from the opposite point of view: we simulate
numerically how accretion dynamics self-consistently create magnetic field
and explore the outflows that result.  Employing the general relativistic
three dimensional MHD code described in De Villiers \& Hawley (2003),  we
have previously investigated the general character of accretion onto black
holes in a series of simulations for which an overview is presented in De
Villiers, Hawley \& Krolik (2003; hereafter Paper~I).  The magnetic field
structures that develop in these accretion flows were described in Hirose
et al. (2004; hereafter Paper~II).  We discussed the resulting outflows
in De Villiers et al. (2005; hereafter Paper~III), and the dynamics of
the inner disk in Krolik, Hawley, \& Hirose (2005; hereafter Paper~IV).
In this paper we return to a more in-depth consideration of the type of
unbound outflows that were discussed in Paper~III.  We will present a
more quantitative and dynamically-oriented analysis of the outflows, and
draw on several new simulations designed to elucidate their properties.
We will focus our attention on three issues in particular: the underlying
principles governing the material component in the outflow; the dynamics
and spin-dependence of the outflow's electromagnetic segment; and the
relationship between the electromagnetic contribution and the class of
ideas placed under the heading of the Blandford-Znajek mechanism.

\section{Overview of Simulations}

\subsection{Qualitative character}

In the present work we solve the equations of ideal MHD in the
metric of a rotating black hole, while assuming that the gas is
adiabatic except in regions of strong compression, where an artificial
bulk viscosity is employed to capture the entropy produced in shocks.
The specific form of the equations
we solve, and the numerical algorithms incorporated into the GRMHD
code, are described in detail in De Villiers \& Hawley (2003).  

In this paper, we present several new simulations, which, like the
simulations reported in Papers I--IV, follow the evolution of an
isolated magnetized gas torus orbiting near a black hole.  The new
simulations differ from the previous ones in part because of several
small technical changes in the code that permitted generation
of new, improved versions of previously-published simulations (e.g.,
raising the cap on the Lorentz factor, saving additional data, extending
the radial grid closer to the horizon for $a/M = 0.9$).  In addition, we
have studied several new high-spin cases: $a/M = 0.93$, 0.95, and 0.99,
and a case with a counter-rotating black hole, $a/M = -0.9$.

\begin{figure}
    \epsscale{0.5}
    \plotone{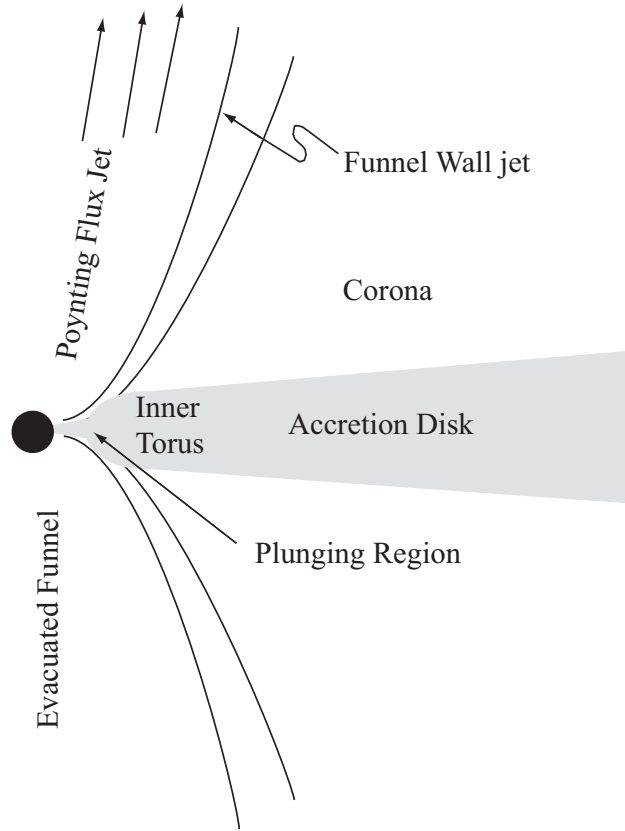}
    \caption{\label{fig:simsum} 
     Schematic illustration of the main dynamical
     features seen in the accretion disk simulations.  
     The jet is defined as those regions of unbound
     positive radial momentum.  
     } 
\end{figure}

The late-time structure of all our simulations contains the same basic
elements pointed out in Paper~I, although their relative importance
and location can change considerably as a function of black hole
spin.  These structures are illustrated in Figure~\ref{fig:simsum},
a schematic of the accretion system as outlined in Papers I through IV.
Along the equator there is a somewhat thickened, wedge-shaped Keplerian
accretion disk.  A net accretion flow is produced within this disk by MHD
turbulence whose ultimate origin is the magneto-rotational instability
(Balbus \& Hawley 1998).  In the disk, the magnetic field is subthermal;
the ratio of the gas to magnetic pressures, i.e., $P_{\rm gas}/P_{\rm mag}
= \beta$, is on average on order 10 to 100.  A short distance outside the
marginally stable orbit around the black hole, the equatorial pressure and
density reach a local maximum in a region we refer to as the inner torus.
The location and vertical thickness of this inner torus can and does
vary as a function of time throughout a simulation.  Inside this local
pressure maximum, the density and pressure drop as the flow accelerates
inward toward the black hole in the plunging region.  The plunging region
typically begins near the innermost marginally stable circular orbit of
the black hole, $r_{ms}$.  The high latitude region above and below the
disk is the corona, a region of hot, magnetized plasma with a magnetic
field whose strength is, on average, near equipartition ($\beta \sim 1$).
Along the spin axis of the black hole there is a centrifugal funnel that
is largely empty of matter, but filled with magnetic field.  This magnetic
field transports energy outward in the form of a Poynting flux.  There is
a region of unbound mass flux at the boundary between the evacuated funnel
and the corona which we refer to as the funnel wall jet.   This funnel
and its boundary are, of course, the center of attention in this paper.


\subsection{Numerical technique}

The GRMHD code we employ for this study solves the equations of ideal
MHD in the metric of a rotating Kerr black hole of mass $M$ in three
spatial dimensions expressed in Boyer-Lindquist coordinates,
$(t,r,\theta,\phi)$.  We use the metric signature
$(-,+,+,+)$.  The determinant of the 4-metric is $g$, $\sqrt{-g} =
\alpha\,\sqrt{\gamma}$, where the lapse function $\alpha=1/\sqrt{-g^{tt}}$,
and $\gamma$ is the determinant of the spatial $3$-metric.  
We follow the usual convention of using Greek
characters to denote full space-time indices and Roman characters for
purely spatial indices.  We use geometrodynamic units where $G = c =
1$; time and distance are in units of the black hole mass, $M$.
The spin of the black hole is specified by $a$, with $a/M \le 1$.

The primitive variables that describe a relativistic test fluid are its
density $\rho$, specific internal energy $\epsilon$, four-velocity
$U^\mu$, isotropic pressure $P$, and magnetic field four-vector $b^\mu$.  The
relativistic enthalpy is $h=1 + \epsilon + P/\rho$.  The pressure is
related to $\rho$ and $\epsilon$ through the equation of state of an
ideal gas, $P=\rho\,\epsilon\,(\Gamma-1)$, where $\Gamma$ is the
adiabatic exponent.  For these simulations we take $\Gamma=5/3$.  The
code evolves auxiliary variables constructed from the relativistic
boost factor $W = \alpha U^t$, namely $D=\rho W$, $E=\rho\epsilon W$,
a four-momentum $S_\mu = (\rho\,h\ + {\|b\|}^2)\,W\,U_\mu$, and a
transport velocity $V^i = U^i/U^t$.  The magnetic field of the fluid is
evolved using the constrained transport magnetic field,
$F_{jk}=[ijk]\,{\cal{B}}^i$, from which the magnetic field four-vector
$\sqrt{4\pi}\,b^\mu = {}^{*}F^{\mu\nu}U_\nu$ is constructed.  The ideal
MHD condition requires $U^\nu F_{\mu\nu} = 0$.

The algorithms employed in this GRMHD code resemble those of the
well-known ZEUS code (Stone \& Norman 1994a,b) used for nonrelativistic
astrophysical MHD simulations.  The evolution is divided into a
transport step, where quantities are advected through the grid by the
velocities $V^i$, and a source step where forces are applied.  We use
the Constrained Transport approach for the induction equation (Evans \&
Hawley 1988) so that  the magnetic field ${{\cal{B}}^i}$ remains
divergence-free.  The evolution timestep is set by the smallest
light-crossing time through a grid zone.

In contrast to some alternative numerical schemes that have been
developed for general relativistic MHD (e.g., Gammie, McKinney \&
T\'oth 2003; Komissarov 2004) our scheme is not fully conservative.
Rather than evolve the stress-energy tensor directly, we work with the
auxiliary variables from which the primitive variables are easily
recovered.  We evolve only the three spatial components of the
four-momentum and use velocity normalization to obtain the $S_t$
component.  The boost factor $W$ is then computed using $S_\mu$, but
because the definition of the inertia used in the code includes $W$, it
is found from the solution to a quartic equation derived from
four-normalization (De Villiers, Staff, \& Ouyed 2005).  The internal
energy is evolved independently.  An artificial viscosity is included
to thermalize some of the kinetic energy lost in shocks, but there is
no mechanism included to capture magnetic energy lost through magnetic
reconnection.  

It is worthwhile to review briefly some of the known strengths and
weaknesses of the GRMHD code.  In tests and applications carried out to
date the code has proven to be fast and robust, capable of carrying out
long-term simulations in three spatial dimensions.  Indeed, as of the time
of the writing of this paper, this is the only code that has been able to
perform long-term (several thousand $M$ in time) three-dimensional GRMHD
simulations.  The code handles both the weak field (i.e., high $\beta =
P_{\rm gas}/P_{\rm mag}$) limit within the accretion disk and the strong
fields ($\beta \ll 1$) found in the evacuated funnel.  Comparisons between
simulations using nonrotating black holes and nonrelativistic simulations
using the ZEUS code show good qualitative agreement within the region
where gravity is well approximated by a pseudo-Newtonian potential.
Our simulations also show qualitative agreement with the axisymmetric
simulations of black hole accretion carried out by McKinney \& Gammie
(2004).

One of the limitations of the code is that there must be a minimum
floor on the fluid density, set in the simulations of this paper at
$10^{-10}$ of the initial density maximum.    This density floor
value is chosen simply to provide a wide dynamic range between the density
within the matter-dominated regions and the region that should be
regarded as the numerical ``vacuum.'' To prevent possible
numerical instability in near-vacuum zones we place a cap on the value
of the boost factor; here $W_{\rm max} = 10$ or $6$.  Because we evolve
an internal energy equation, the thermodynamics are only approximate.
Some amount of energy that would otherwise be thermalized (e.g., in
magnetic reconnection) is lost.  As we have no explicit cooling, the
disks grow hot as they accrete, but not as hot as they would be with
complete energy conservation.  Conversely, of course, they are
not as cool as they might be if radiation losses were taken into account.

Because we use Boyer-Lindquist coordinates, the horizon represents a
coordinate singularity, and our inner grid boundary, $r_{\rm in}$, must
lie outside this point.  The metric terms increase rapidly near the
inner boundary, making accurate numerical calculations there more
difficult.  In practice we select an inner radius such that the Kerr
metric term $\Delta = r^2-2Mr +(a/M)^2$ is approximately $\sim
0.1$--$0.02$.  As a result, the spacetime rotation frequency at
$r_{\rm in}$ is less than that of the horizon.  The inner boundary uses
zero gradient conditions on the fundamental variables, and the radial
velocity and momentum at $r_{\rm in}$ are not allowed to become
positive.

Even apart from the coordinate singularities at the horizon in
Boyer-Lindquist coordinates, it
is well known that spherical coordinates have a singularity along the
axis.  As a practical matter, the closer one is to the $\theta = 0$ or
$\pi$ axis, the more narrow and wedge-like the $\phi$ zones become,
leading to a substantial reduction in the Courant-limited timestep
among other problems.   To avoid this, we set the $\theta$ boundary not
at the axis but at a small angle off the axis and employ reflecting
boundary conditions.  This means that with our grid we cannot
follow a flow passing over the north or south pole of the black hole.

These conditions imposed on the computational grid will, of course,
affect the results.  The ``torque'' applied by the rotating spacetime
will be limited to that provided exterior to the inner boundary,
$r_{in}$, and there will be no jet power along the cutout on the axis.
The numbers reported for the models should not be regarded
as quantitatively
predictive of the jet power a black hole with a specific $a/M$ value
would have in Nature.  Rather we are interested in the relative values of
quantities across the full complement of the simulations and what that
can mean for the role of a rotating hole in the combined
accretion plus jet production process.

It is important to keep these code limitations in perspective.
As no numerical scheme is perfect, all simulators must make choices.
There are often alternatives that can address one specific limitation,
but introduce new limitations of their own.  For example, the problems
associated with a spherical mesh can be eliminated through the use of
a Cartesian grid, but then angular momentum is no longer a directly
conserved fundamental variable and maintaining a simple orbit becomes a
matter of carefully balancing forces acting on distinct linear momentum
components.  Diffusion error associated with motion across the grid can
then be a significant source of nonphysical viscosity in an accretion
disk.  When evolving a total energy equation, the internal energy is
defined as what remains after all other independently evolved  energies
(kinetic, magnetic) have been subtracted off.  If the internal energy is
a small fraction of the total energy (as it is whenever $h/r \ll 1$),
its evolution can be dominated by truncation errors in the kinetic
and magnetic components.  In relativity, the derivation of primitive
variables from conserved quantities is difficult and time-consuming.

At this stage the only practical approach is to simulate well-chosen
problems with a variety of numerical codes and to evaluate the
outcomes both in comparison to other results and with
self-consistency checks, keeping a specific code's limitations firmly
in mind.

\subsection{Problem specifications}

For this paper we have computed a new set of models in a spherical
quadrant computational domain. The inner radial boundary, denoted
$r_{\rm in}$, is placed just outside the event horizon (the exact
amount varying from case to case), and the outer radial boundary is at
$r_{\rm out} =120M$.  The azimuthal angle ranges over $0 \leq \phi \leq
\pi/2$ and the polar angle over $0.045\pi \leq \theta \leq 0.955\pi$.
The radial boundary conditions are pure outflow, the azimuthal boundary
conditions are periodic, and the boundary conditions near the polar
axis are reflecting.  The grid consists of 192 radial cells spaced
using a hyperbolic cosine function in order to maximize the resolution
near the inner boundary,  192 polar angle cells, with an exponential
grid spacing function that concentrates zones near the equator, and 64
evenly-spaced cells in azimuthal angle.  This resolution is close to
the maximum feasible at this time for computing an ensemble of three
dimensional simulations.

These models are of the same form as the KD models described in
Paper~I.  Since those simulations were run, we have added new
run-time diagnostics and
have studied a greater range of black hole spin parameters: $a/M =
-0.9$, 0., 0.5, 0.9, 0.93, 0,95, and 0.99.  Note that we have included
one case with a black hole that is counter-rotating with respect to the
disk.  As in Paper~I, the initial conditions consist of an orbiting
isolated gas torus with a pressure maximum located at $r \approx 25M$,
and a slightly sub-Keplerian initial distribution of angular momentum
throughout.  We construct an initial torus so that its inner edge is
located at $r=15 M$ regardless of the black hole spin rate.  The
initial magnetic field consists of loops of weak poloidal field lying
along isodensity surfaces within the torus.  The strength of the
magnetic field is chosen so that the ratio of the total magnetic
pressure to the initial total gas pressure in the torus is 0.01 ($\beta
= 100$).  Table~\ref{table:sims} lists the models along with some
reference values such as the radius of the marginally stable orbit,
$r_{\rm ms}$, and the frame-dragging rate at the inner grid boundary,
$\omega(r_{\rm in})$.

\section{Global Measures of the Outflow}

The strength of the outflow may be characterized by how much mass,
linear momentum, angular momentum, and energy it carries to large
distance.  We compute these measures through shell-integrated fluxes
that are calculated using 
\begin{equation}\label{avgdef} 
   \langle{\cal F}\rangle(r,t) =
      \int\int{{\cal Q}(r) \,\sqrt{-g}\, d \theta\,d \phi},
\end{equation} 
for quantity ${\cal Q}$ at radius $r$, where the bounds
of integration range over the full $\theta$ and $\phi$ computational
domains.  These fluxes are computed and stored every $M$ in time, and
integrating the flux over time yields the total outflow or accretion
rate.  Whenever we speak of mean properties of the outflow, we require
a criterion to distinguish those cells participating in the outflow
from those in the accretion flow.  Our definition of ``outflow" is any
cell with $-h U_t > 1$ (i.e., that is unbound) and with $U^r > 0$
(i.e., moving outward).   This criterion works well for these
simulatons, as only the outflow near the axis (the ``jet'' outflow) is
unbound; the backflow from the disk itself remains bound.  In practice
we divide the shell integrals into two parts, one for bound and one for
unbound flow.  We will indicate the shell integrals specific to either
the outflow or the accretion flow by multiplying the integrand by
either ${\cal W}$ or $1-{\cal W}$, where ${\cal W}$ is a function whose
value is unity in outflow cells and zero otherwise.

In the present work we are concerned with the strength of the jet
compared to what is accreted into the hole.
We would like to characterize the outflows in terms of the accretion rate,
but this can be ambiguous because of the scale-free nature of these
simulations (code values of density are arbitrary), and the fact that
the total mass of the initial torus is a function of black hole spin.
We proceed by computing the total mass, angular momentum and energy
(in code units) that leave the grid going into the black hole; these
are the ``hole'' values.  Second we compute the total {\it unbound}
mass, angular momentum and energy that pass through $r=100M$; these are
the ``jet'' values.  To focus on the quasi-steady state portion of the
simulation, we compute these totals from $t=3000$ to $10^4 M$.  
These integrated quantities
are listed in Tables~\ref{table:mflux},
\ref{table:lflux} and \ref{table:eflux}, along with 
measures of the outflow efficiency.  For the
case of the angular momentum and energy fluxes, we further distinguish
between that carried by the matter (superscript $m$) and that carried
by the electromagnetic fields (superscript $em$).
The efficiencies are defined as
\begin{eqnarray}
\eta_{m} &\equiv& \left(E^{m}_{\rm jet} - M_{\rm jet}\right)/M_{\rm hole}\\
\eta_{em} &\equiv& E^{em}_{\rm jet}/M_{\rm hole},
\end{eqnarray}
where $M_{\rm hole}$ is the total rest-mass accreted
into the black hole, while $M_{\rm jet}$ and $E_{\rm jet}$ are
the rest-mass and
energy carried off through the $r=100M$ surface.  Rest-mass energy is
subtracted from the matter portion so that the efficiency refers only
to ``usable" energy.  In addition to relative efficiency,
it is also useful to 
compared to the standard Novikov-Thorne (Novikov \& Thorne 1973)
thin disk efficiency,
which is the binding energy of the innermost stable circular orbit.
These efficiencies are also listed in Table~\ref{table:eflux}.

Before discussing these numbers, we must first express a caution:
the fluxes are, in general, not constant in time or radius.  Total
(i.e., electromagnetic plus matter) quantities can depend on radius,
even in the long-term average.  Even where the total is constant,
electromagnetic fields and matter can, and do, exchange energy and
momentum.  In most cases, for example, the EM outflow rates reach a
peak at several tens of gravitational radii and decline slowly from
there to the outer boundary.  In contrast, the magnitude of the
time-averaged rest-mass expulsion rate increases monotonically toward
the outside, although its rate of increase decreases with growing
distance from the center.  In a strict sense, the outflow numbers apply
only at our fiducial outflow radius $r=100M$.  Consequently,
the numbers we give in the tables should be regarded as useful for
comparing relative changes between simulations, and
to study how outflow properties depend on black hole spin,
not as providing some sort of absolute measure of jet power.  

Examining the table of mass and energy flows, the most obvious
conclusion is that for rapidly-rotating black holes the jet efflux is
not insignificant compared to what is accreted.  In other words, the
jet plays a noticeable role in the global energy budget of accreting
black holes.  The mass outflow is all within the funnel wall jet,
whereas the Poynting flux outflow is primarily within the funnel
itself.  The ratio of jet mass to accretion mass is generally a few
percent, although it rises steeply from $a/M = 0.95$ to $a/M = 0.99$.
Moreover, although the trend in these simulation data is not monotonic,
there is a general tendency for the expelled energy to increase sharply
with increasing black hole spin.  The outflows are almost nil in the
Schwarzschild case, weak when $a/M = 0.5$, and become substantial when
$|a/M| \geq 0.9$.

When $|a/M| \geq 0.9$, the total jet efficiency is generally several tens
of percent, with the matter portion somewhat larger than the
electromagnetic part.   The retrograde case is less efficient in
driving a jet than the prograde case with the same magnitude of spin,
but only by a factor of two.  Interestingly, in otherwise similar
simulations most of whose results closely resemble ours, McKinney
(2005) remarks that he finds the retrograde jet efficiency to be smaller by
a factor of ten.

The ratio of angular momentum in the unbound outflow to that deposited
in the black hole is also a very strong function of black hole spin.
Only a few tenths of a percent in the Schwarzschild case, it rises to
$\simeq 25\%$ when $a/M = 0.9$.  In the highest spin case reported here
($a/M = 0.99$), the {\it net} angular momentum accreted is extremely small
because the rate at which the black hole spins down via electromagnetic
torques nearly matches the rate at which it acquires angular momentum
by accreting matter.  Consequently, the angular momentum expelled in
the outflow is an order of magnitude {\it greater} than the net amount
captured.  In all cases, the electromagnetic portion of the angular
momentum carried outward is comparable to the matter portion carried
in the funnel wall jet.

Note in particular that the electromagnetic angular momentum carried
outward in the retrograde case has the opposite sign relative to
the accreting matter's orbital angular momentum.  One consequence of
this is that the net angular momentum outflow in this case
is very small because the electromagnetic part very nearly cancels the
matter part.  This fact is also an indication (that will be confirmed
in the following section) that the electromagnetic outflow is directly
tied to the black hole's rotation.

\section{Electromagnetic Outflow: the Funnel Interior}

\subsection{Character of magnetic field in the funnel}

In the initial state there is no magnetic field at all within what
eventually becomes the outflow region.  All of the simulations
presented here rapidly develop a large scale field within the funnel,
which in turn leads to the formation of the Poynting flux jet.  This is
one of the significant results of simulations of this type (see also
Paper I and II, as well as McKinney \& Gammie 2004),
namely that a large scale, jet-driving magnetic field can develop
naturally as the result of magnetically controlled accretion.

\begin{figure}
\centerline{\psfig{file=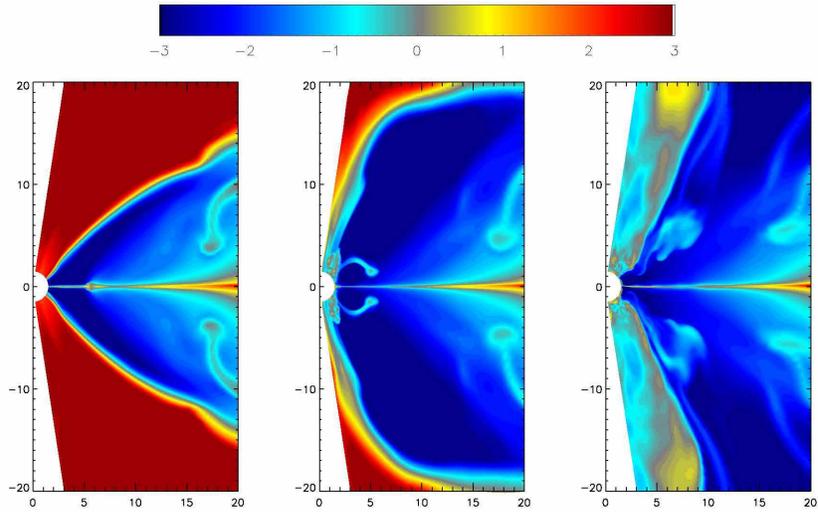,angle=90,width=5.5in}}
\caption{Ratio of the azimuthally-averaged gas to magnetic pressure 
$\beta$ at $t=560M$, $640$M, and $720M$ in KDPg.  The color contours
are in a logarithmic scale:  dark red is gas pressure dominated 
($\beta = 10^3$) and  dark blue is magnetic field dominated ($\beta = 0.001$). 
\label{fig:fieldeject}}
\end{figure}

This funnel field formation  process may conveniently be visualized in
terms of three steps.  These are illustrated in the panels of
Fig.~\ref{fig:fieldeject}, which show the evolution of the $\beta$
parameter in the inner region of the inflow in run KDPg.  First, as the
magnetorotational instability (MRI) develops in the disk, matter begins
to move inward.  The magnetic field is stretched radially, and this
radial field is sheared into an increasingly strong toroidal field.
An extremely thin precursor of matter moves quickly inward in the
equatorial plane, surrounded by a thicker envelope of
magnetically-dominated ($\beta \sim 0.1$) matter.  The matter density
above and below the equatorial accretion flow is very low, in part
because of the initial conditions.  However, the expectation of low
density above and below the plunging inflow is likely to be general
because it is the centrifugally-excluded zone.  When the accretion flow
reaches the black hole, matter can drain off the field lines.  The
vertical component of $\nabla B^2$ is then great enough to drive a
vertical expansion of the field (center panel in
Fig.~\ref{fig:fieldeject}).  A ``magnetic tower" forms, as has
previously been predicted on analytic grounds (Lynden-Bell 2003) and
seen in other simulations (Kato, Mineshige, \& Shibata 2004).  It lasts
for a time $\sim 100$--$200M$.


The steady-state structure is created in the third step.  Because the
top of the ``magnetic tower" expands more rapidly than its base, the
toroidal wrapping of field lines stretches into a less tightly-wound
helix.   To stretch the field from the ``tower" region $\sim 20M$ from
the plane all the way to the outer boundary requires only $\sim 100M$
in time because the motion is relativistic.  When this happens in a
Schwarzschild metric, the field lines end up very nearly radial; with
increasing black hole spin, the field lines wind more tightly (Paper~II).
That is, it is only the intrinsically general relativistic effect of
frame-dragging that in the steady state imparts a net helical twist to
the field far from the black hole.  Any remnant of the ``tower" that
persists in the long-run is not due to MHD accretion dynamics; it is,
rather, an imprint of the rotating metric.

As discussed in Paper~II, the interior of the outflow funnel is
magnetically-dominated in the sense that $||b||^2/(\rho h) \gg 1$.
Values of this ratio in the range $\sim 10$--$10^3$ are common.
However, we stress that the particular value of $\rho h$ found in the
funnel is not well-determined in these simulations.   The density
reaches the floor value at late time in much of the funnel.  This occurs
for all values of the floor that have been tested; clearly the funnel
is being evacuated.  Except for
some low density initial background material, the matter on the grid
all has angular momentum, and this prevents it from entering the axial
funnel.  Thus the finding that the funnel should be evacuated is
expected, and our results should simply be taken as saying that the
magnetic energy density is likely to be at least several orders of
magnitude greater than $\rho h$.

\begin{figure}
\centerline{\psfig{file=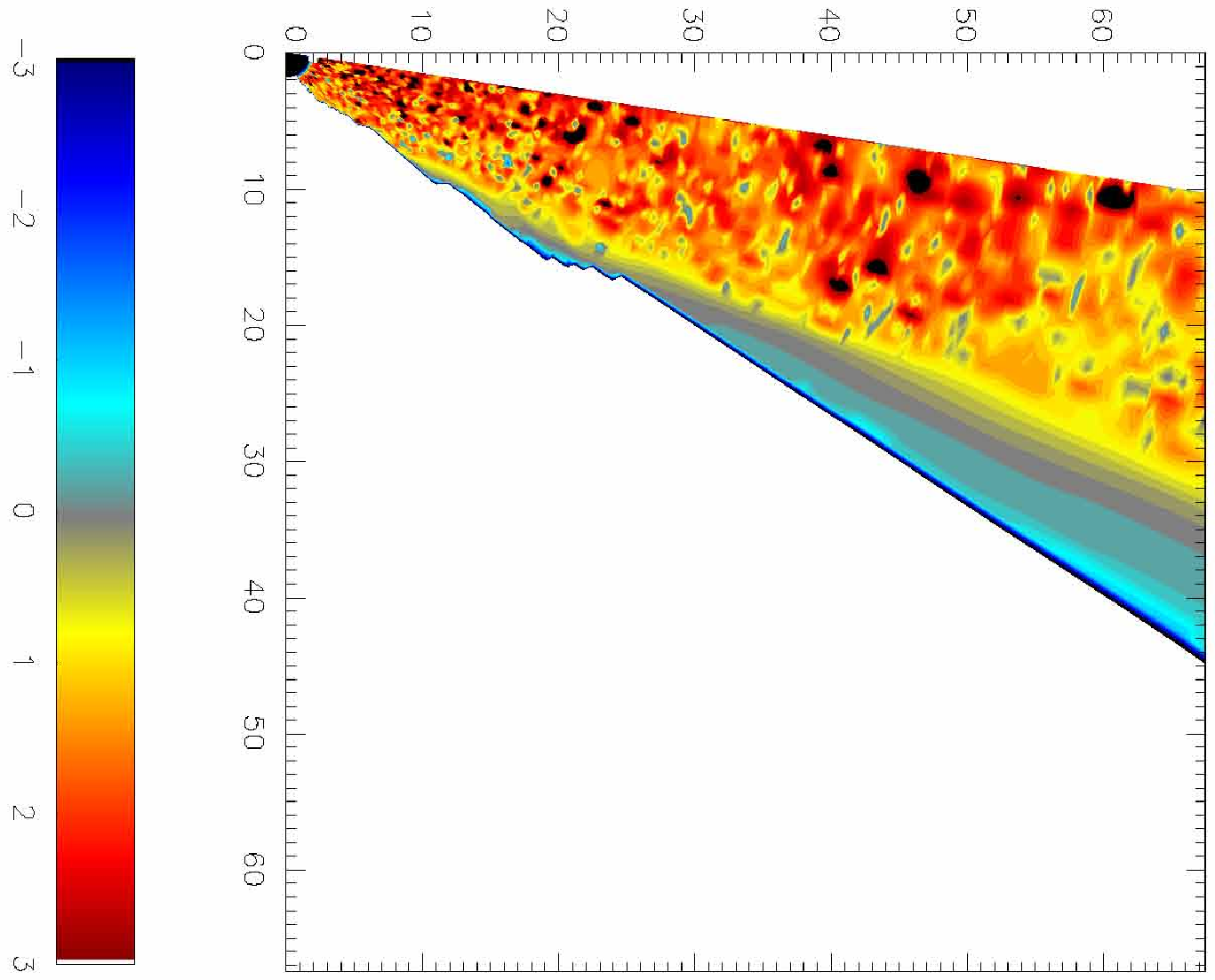,angle=90,width=2.4in}
\psfig{file=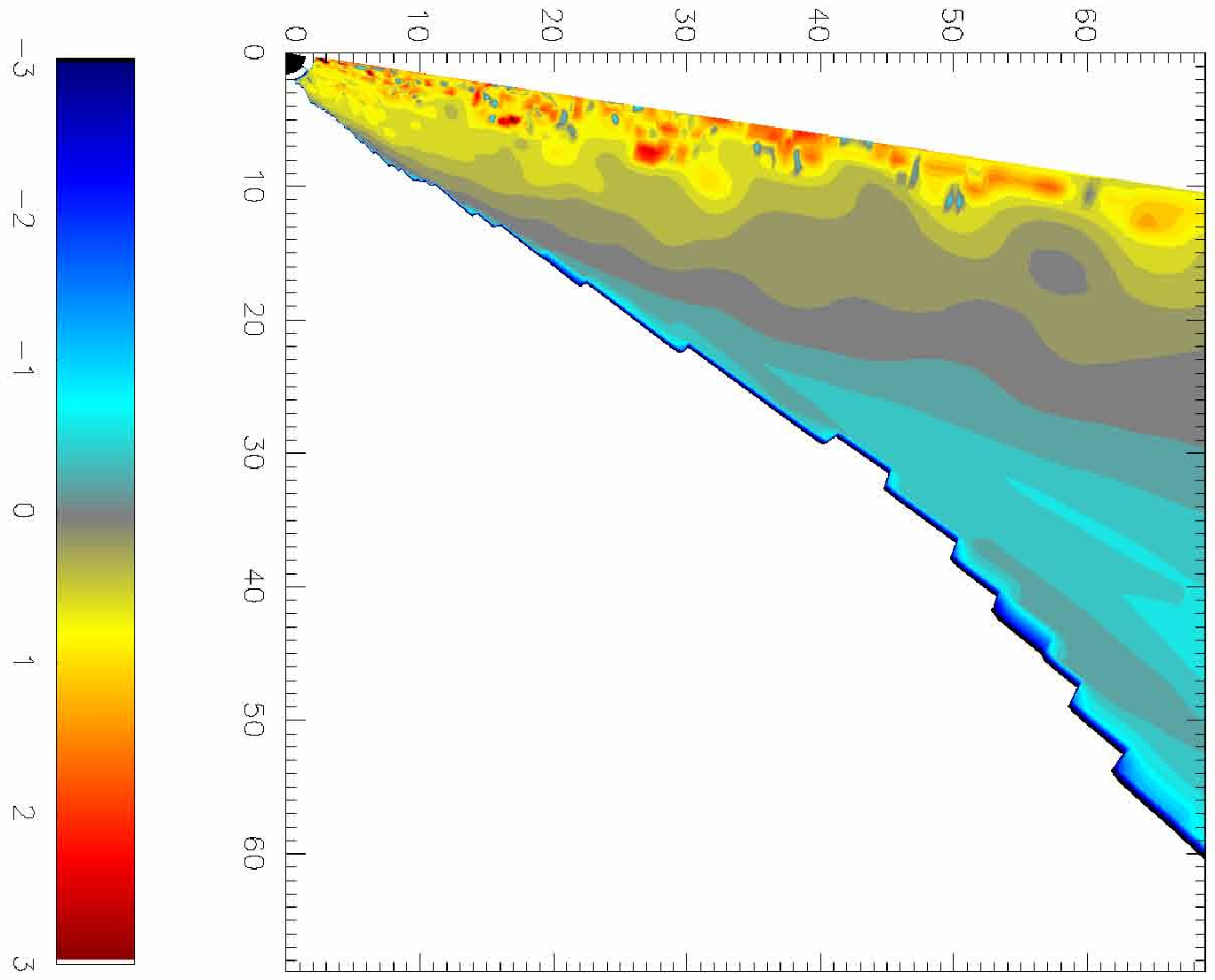,angle=90,width=2.4in}
\psfig{file=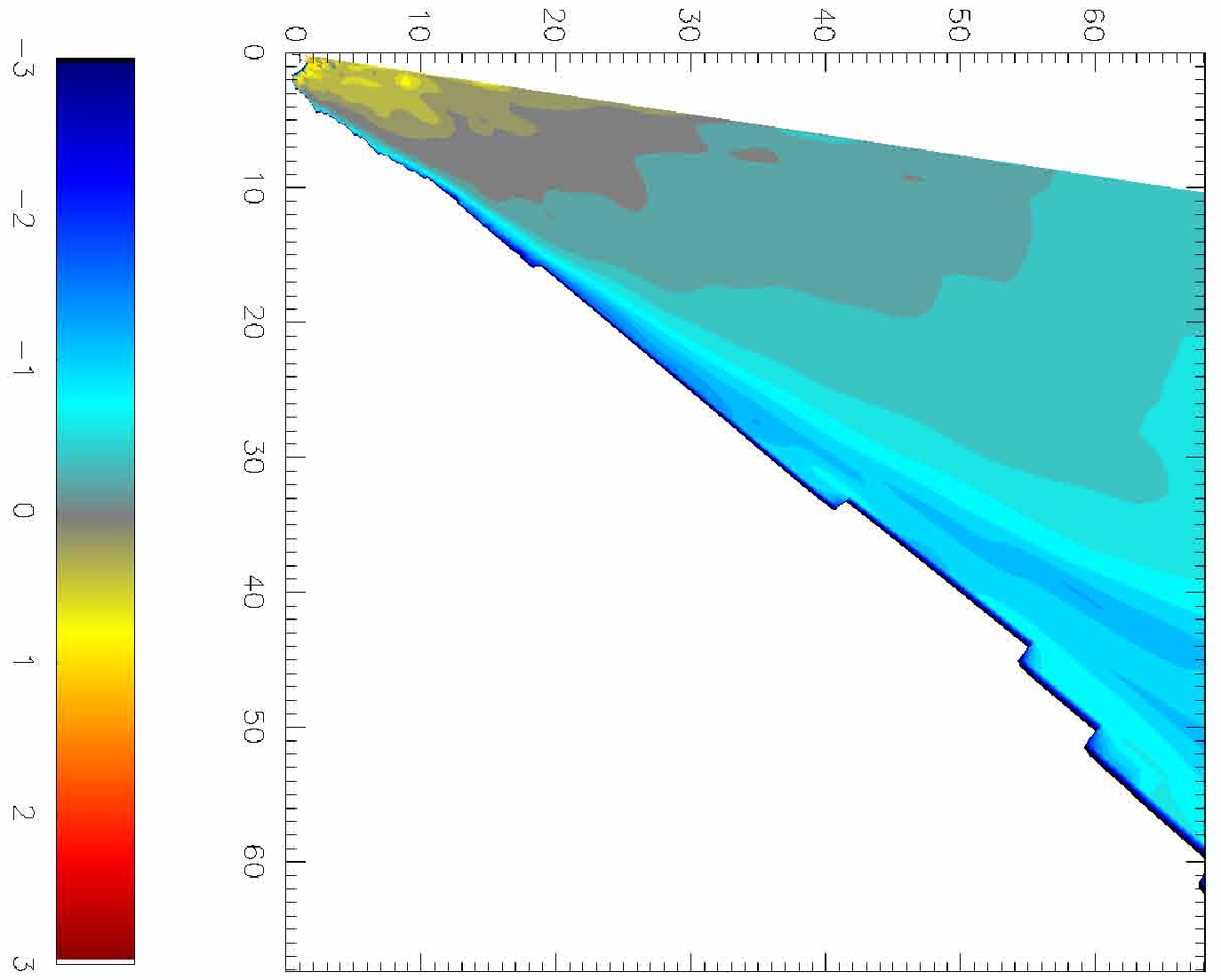,angle=90,width=2.4in}}
\caption{Azimuthal mean of the magnetic radial to azimuthal component ratio
at a late time in the $a/M = 0$, $a/M=0.5$, and
$a/M=0.9$ simulations, from left to right.  The contours are logarithmic.
\label{fig:brbphratio}}
\end{figure}

The steady-state funnel field is illustrated in
Figure~\ref{fig:brbphratio}, which shows the ratio of radial to
azimuthal components of the magnetic field in the outflow at a late
time in three of the simulations.  Because the $g_{t\phi}$ term in the
metric becomes small when $r \gg M$, and the focus here is on behavior
at large distances, we have approximated this ratio as if the metric
were purely diagonal, i.e., we have plotted the ratio $({\cal B}^r/{\cal
B}^{\phi})\left({g_{rr}/g_{\phi\phi}}\right)^{1/2}$.  
As the figure shows, the field
is very purely radial when the black hole does not spin, but grows
progressively more toroidal with increasing black hole rotation rate.
For intermediate values of spin, the field is more nearly radial near
the polar axis, more toroidal at larger angles.

\clearpage

\clearpage

If this flow were time-steady and axi-symmetric,
it would make sense to discuss the rate at which the poloidal portion
of the field rotates around the axis.  Under those conditions the fieldline
rotation rate is defined as 
$\omega = V^\phi - {\cal B}^\phi V^r/{\cal B}^r$
or equivalently as
$\omega = V^\phi - {\cal B}^\phi V^\theta/{\cal B}^\theta$.
Here, because the situation is
neither axi-symmetric nor stationary, the concept of fieldline rotation
rate is not so clear.  Nonetheless, in order to provide some intuition
by comparison to a more familiar context, we construct a local and
instantaneous definition of fieldline rotation rate:
\begin{equation}\label{eq:fieldlinerot}
\omega \equiv V^\phi - 
 {\cal B}^\phi\frac{V^r{\cal B}^r g_{rr}
   +V^\theta{\cal B}^\theta g_{\theta\theta}}
   {({\cal B}^r)^2 g_{rr} + ({\cal B}^\theta)^2 g_{\theta\theta}}.
\end{equation}
The second term describes how fast matter slips along the toroidal
field relative to the rotation of the poloidal part.  In the time-steady
axi-symmetric limit, equation~\ref{eq:fieldlinerot} reduces to the standard
definition of fieldline rotation (as found, e.g., in Phinney 1983).

\begin{figure}
\psfig{file=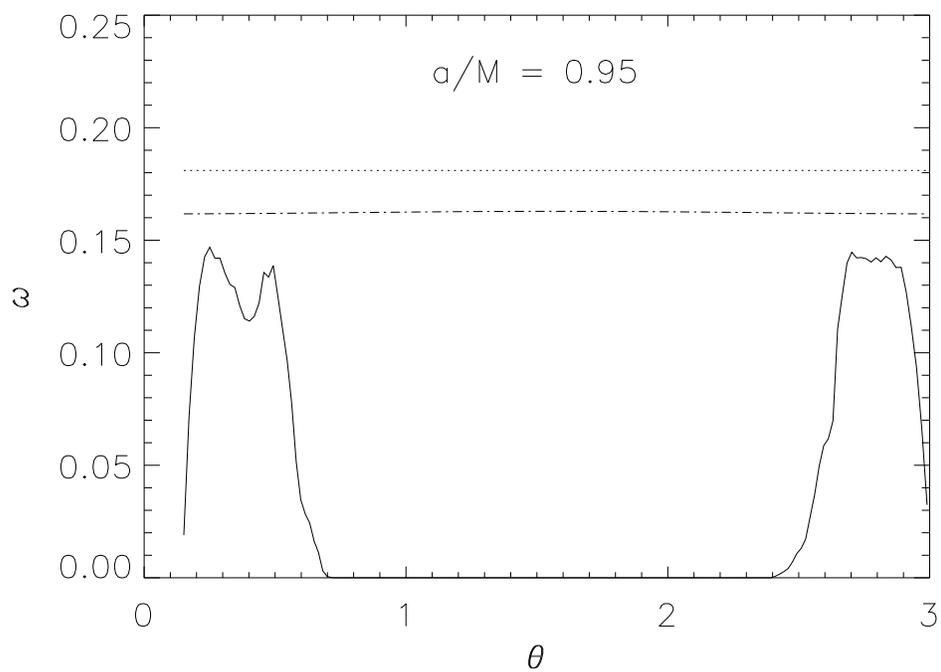,angle=90,width=5.5in}
\caption{Fieldline rotation (as defined in text) in the outflow
for the simulation
with $a/M = 0.95$.  The horizontal lines show half the rotation rate
of the black hole (dotted) and half the rotation rate of the inner
boundary (dash-dot). \label{fig:fieldrot}}
\end{figure}

In Figure~\ref{fig:fieldrot} we show how this definition of the rotation
rate compares to specific predictions made in the context of
the Blandford-Znajek model.  When, for example, the fieldlines are
radial, it is expected that the rotation rate close to the polar
axis should be half the black hole rotation rate
(Blandford \& Znajek 1977; Phinney 1983).  To make this comparison,
we must average over both position along individual fieldlines and
over time.  For the former average, we make use of the fact that 
the flow is nearly conical at large radius, and define
\begin{equation}
\langle \omega \rangle (t,\theta) \equiv
(2/\pi)\int_0^{\pi/2} \, d\phi \, \int_{r_{\rm in}}^{r_{\rm out}} \, dr \,
            \omega (t,r,\theta,\phi) ,
\end{equation}
where we choose $r_{\rm in} = r_{\rm out}/2$ in order to restrict the average
to the nearly conical region.  In practise, although we see substantial
fluctuations in $\omega$ as a function of radius at fixed $\theta$,
we see little in the way of systematic trends with radius in this outer
region, so the exact range of averaging should not make much difference.


The details differ slightly as a function of time within any one
simulation and also from simulation to simulation.  For example, the
behavior of the field line rotation at the axis boundary varies,
sometimes dropping off as in Figure~\ref{fig:fieldrot} and sometimes
increasing. (Recall that the axis has a reflecting boundary condition
applied at $\theta = 0.045 \pi$.)  Within the main part of the funnel,
however, things are more consistent from model to model.  The fieldline
rotation rate is close to, but generally $\simeq 20$--$25\%$ smaller
than, half the spacetime rotation rate of the inner boundary (see
Table~\ref{table:sims}).  We use this criterion rather than half the
rotation rate of the black hole itself because the latter quantity
plays no direct role in our simulation.  The difference between the
classical prediction and the observed rotation rate in our results is
larger than that observed in the two-dimensional simulations of
McKinney \& Gammie (2004).  They found that, within the force-free part
of the outflow, the rotation rate was about $10\%$ less than half
the black hole rotation rate.

We further find that the rotation rate of the funnel fieldlines is
always in the black hole's sense; that is, in the retrograde case, the
fieldlines rotate in the same direction as the black hole, but in the
opposite direction of the accreting matter.  Specifically, defining the
black hole spin as in the positive direction, when the disk and the
black hole rotate in opposite senses, the rotation of the funnel-wall
matter is backward, and, at large radius, of smaller magnitude than the
fieldline rotation.  This contrast underlines the fact that the
fieldlines' rotation is fundamentally a result of
frame-dragging in the ergosphere.  Interestingly, the mean rotation
rate in the retrograde case is actually considerably closer to half the
rotation at the inner boundary than it is in the prograde cases.

Finally, although we have hopes of finding large-scale field-generation
processes that are completely general, we note that the funnel field
formation seen here does depend in certain ways upon the problem
configuration.   First, the crucial ingredient is the presence of radial
field extending through the plunging region as material first accretes
into the hole.  The polarity of this field, and of the subsequent funnel
field, is a consequence of the initial field configuration, namely
poloidal loops with one sign of the radial field above the equator and
the opposite sign below.  A simulation with a purely toroidal initial
field (discussed in Paper III) is a counter-example that does not form
an organized funnel field.  Second, the initial plunging inflow is
surrounded by a near-vacuum which offers no opposition to the formation
of the magnetic tower.  Third, the accretion flow is relatively hot,
with $H/R \sim 0.15$.  It is possible that thinner, colder disks would
not generate sufficiently strong fields to create the magnetic tower;
additional simulations will be necessary in order to explore the degree
of generality of this large-scale field-generation mechanism.

\subsection{Angular distribution of Poynting flux}

In the time-steady, axi-symmetric limit in which it is possible to
define clearly a fieldline rotation rate, the Poynting flux on the
inner boundary can be described in terms of the magnetic field strength
and the fieldline rotation rate:
\begin{equation}\label{eqn:bzpoynting}
{\cal F} = \omega r_{\rm in} {\cal B}^r {\cal B}^\phi (\Delta/\Sigma^2)
                           \sin^2 \theta,
\end{equation}
where $\Sigma = r^2 + (a/M)^2\cos^2\theta$ and $\Delta = r^2 - 2Mr + (a/M)^2$
are the standard functions appearing in the Boyer-Lindquist metric
(Blandford \& Znajek 1977).  Although $\Delta \rightarrow 0$ as the
event horizon is approached, the magnetic field strength diverges in
an appropriate way to make ${\cal F}$ continuous.
In the specific circumstances envisaged by Blandford \& Znajek (1977),
the polar angle dependence of the magnetic field intensity follows
from the condition
of satisfying a force-free equilibrium for a field of specified multipolar
character.  They predict that for a split monopolar field it should diminish
by about 20\% from the axis to the midplane.  In addition, they predict
that the fieldline rotation rate should be close to 1/2 the black hole
rotation rate almost independent of polar angle, so the overall polar
angle dependence of the Poynting flux should be almost $\propto \sin^2\theta$.

\begin{figure}
\centerline{\psfig{file=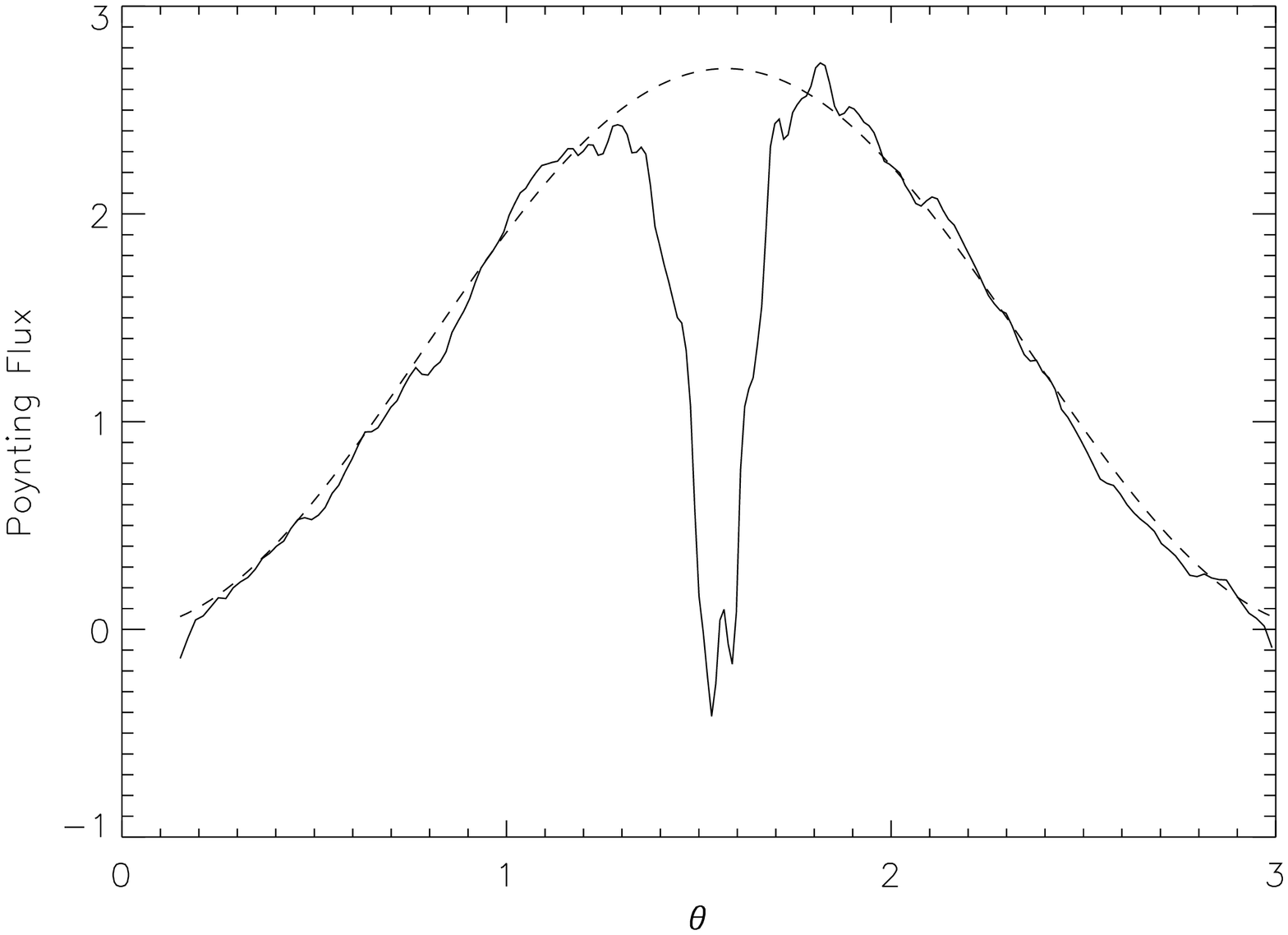,angle=90,width=2.75in}
     \psfig{file=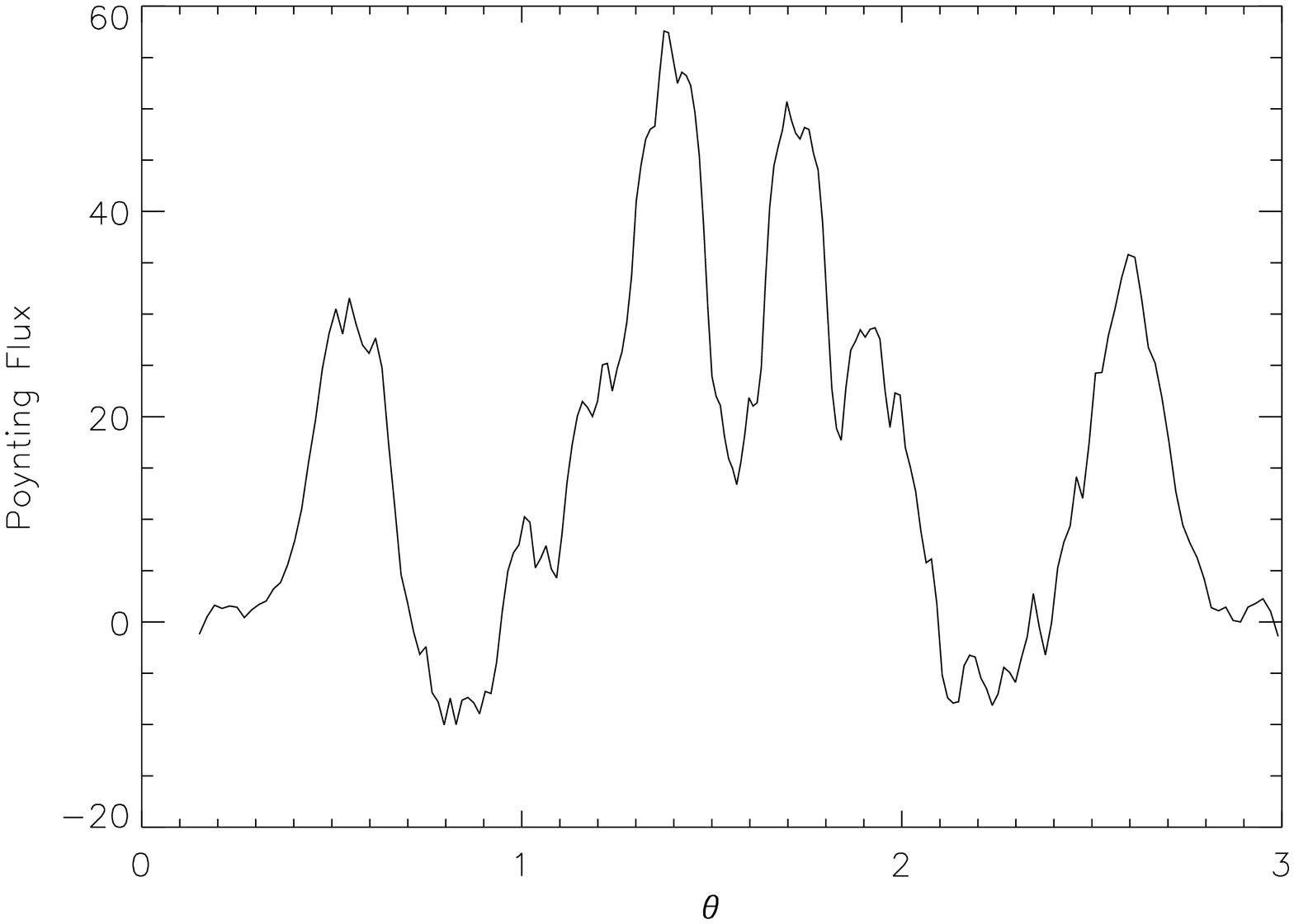,angle=90,width=2.75in}}
\caption{Time-averaged Poynting flux on the inner boundary in code units
times $10^6$.  Left panel is KDIb, averaged over 26 snapshots 
equally spaced from
$t=8000M$ to $10000M$.  The dashed line shows a simple fit to
$\sin^2\theta$.  Right panel is KDH, averaged over 27 snapshots
from $t=7920M$ to $10000M$. \label{fig:Poynting}}
\end{figure}

In our simulation, we find that the magnetic field sufficiently
dominates the matter inside the funnel that a force-free configuration
might be expected on average (although certainly not obtained
instantaneously, as the fluctuations are large), while the detailed
dynamics of the coupled accretion and jet flows determine the geometry
of the field.  In the simulations with $|a/M| \leq 0.9$, the
time-averaged distribution of ${\cal F}$ (calculated directly from the
electromagnetic contributions to $T^r_t$, not from
equation~\ref{eqn:bzpoynting}) does indeed vary with polar angle very
nearly as $\sin^2 \theta$ (Fig.~\ref{fig:Poynting}).  However, this
equilibrium structure is not necessarily realized at any given time,
for there are tens of percent fluctuations around the mean value.   In
the larger spin cases, these fluctuations are larger in magnitude.
Even after averaging over some 27 snapshots, the polar angle dependence
of the Poynting flux in the $a/M = 0.95$ case is still quite rough.  It
is also noteworthy that, although the Poynting flux goes inward in the
midplane accretion flow at lower spins, for $a/M \geq 0.95$, it is
outward almost everywhere, even within the equatorial accretion flow.
The overall pattern seen in the $a/M = 0.95$ case---a rise from the
polar axis to mid-latitudes, a sharp drop at an angle $\simeq \pi/4$
from the pole, and irregularly-distributed but intense outgoing flux
nearer the midplane---was also seen in the two dimensional 
$a/M = 0.938$ simulation of McKinney \& Gammie (2004).


\section{Mass Outflow}

\subsection{The centrifugal barrier}

\begin{figure}
\psfig{file=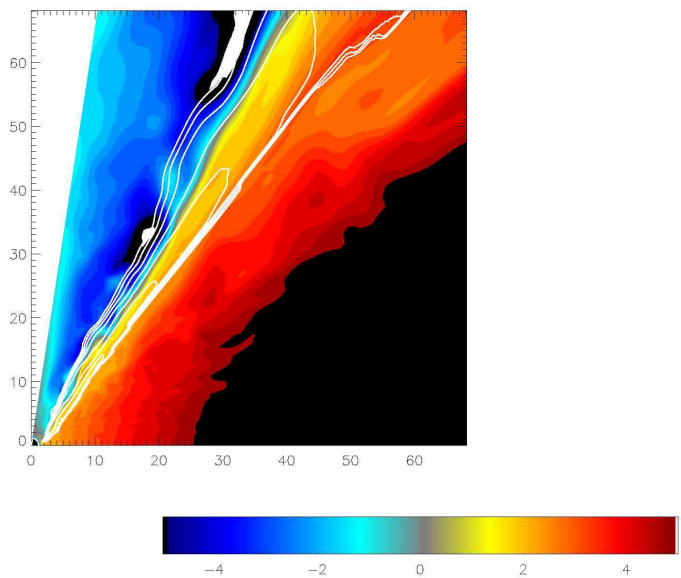,angle=90,width=5.5in}
\caption{Specific angular momentum and mass flux at late time in the
KDPg simulation.  Color contours show $\langle U_\phi \rangle_\rho$.  
Superimposed
line contours are logarithm of the mass flux in the outflow.  Peak mass
flux follows very closely along the $\langle U_\phi \rangle_\rho \simeq 2$
contour.\label{fig:angmomdist}}
\end{figure}

We confirm by direct examination of the complete data files at late
time that the mass outflow in the jets is confined to the funnel wall
component.  In fact within the funnel at the radius for which the
outflow fluxes are measured, $r=100M$, the density inside the funnel
has dropped to the floor value; matter is, of course, excluded from
the funnel if it has significant angular momentum.
The mass-weighted mean angular momentum in the funnel-wall outflow
can be computed as
\begin{equation}
\langle U_\phi \rangle_\rho \equiv 
    \int \, dV \, \sqrt{-g} \rho U_\phi {\cal W}/
        \int \, dV \sqrt{-g} \rho {\cal W}.
\end{equation}
This quantity is always comparable to the angular momentum
of Keplerian orbits in the innermost part of the disk.  For example,
in KDPg, $\langle U_\phi \rangle_\rho$ rises from $\simeq 1$ at
the base of the outflow ($r \simeq 3M$) to $\simeq 2$ at the
outer boundary (see Fig.~\ref{fig:angmomdist}); by comparison,
$U_\phi (r_{\rm ms}) \equiv l_{\rm ms} = 2.1$.
At higher spin, $\langle U_\phi \rangle_\rho$ can rise as high as
$\simeq 3$, or roughly twice $l_{\rm ms}$.


\begin{figure}
\centerline{\psfig{file=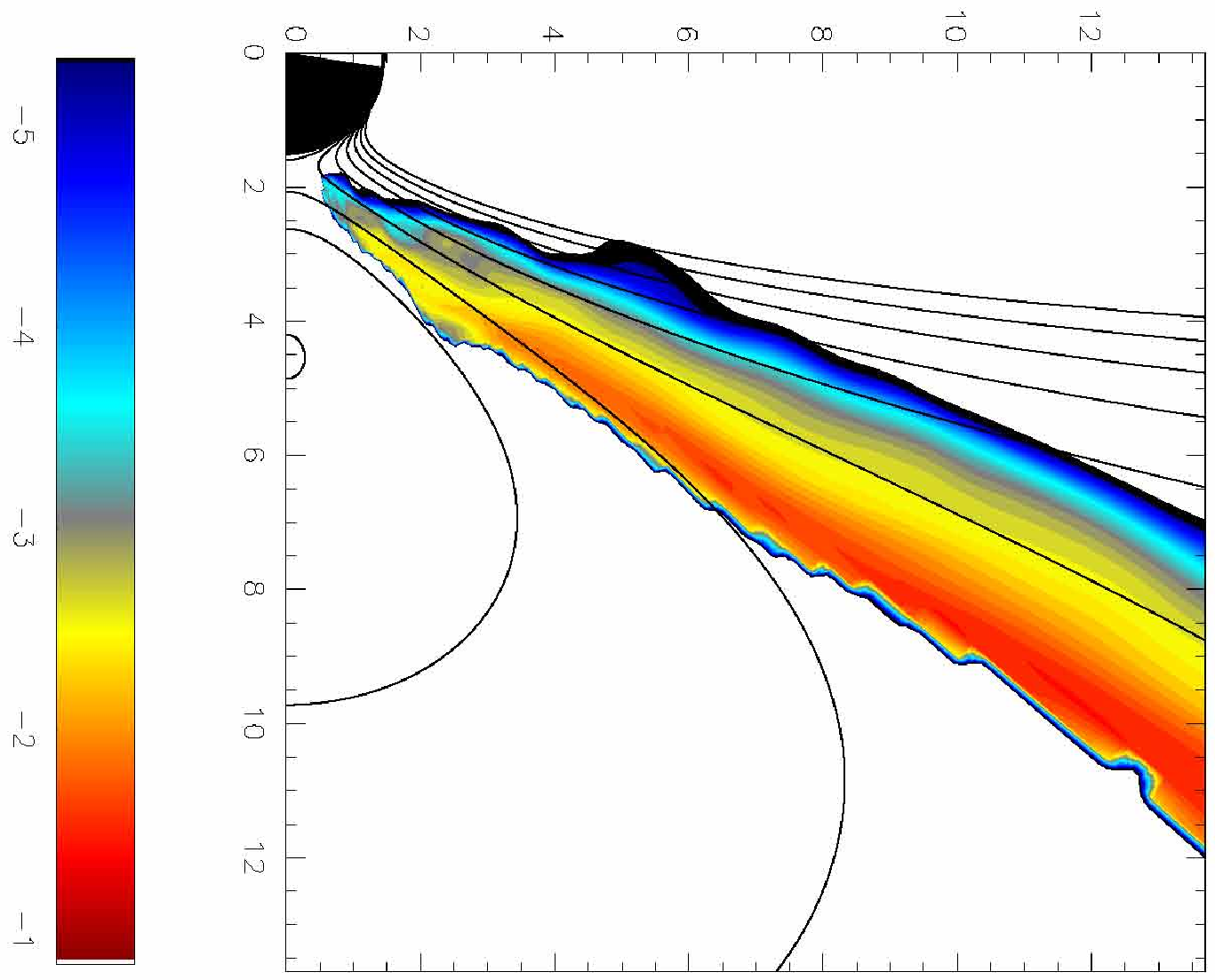,angle=90,width=2.75in}
            \psfig{file=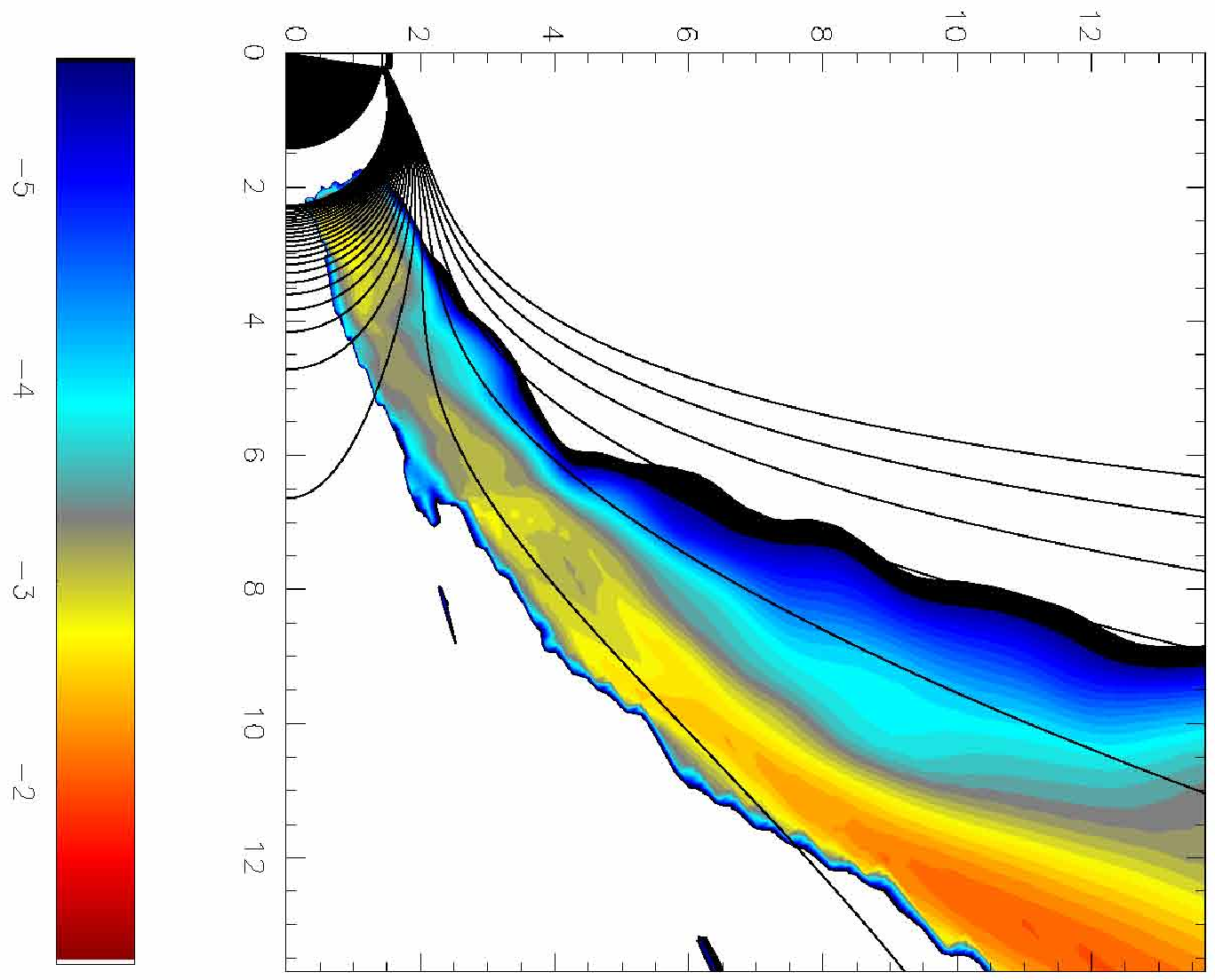,angle=90,width=2.75in}}
\caption{Mass outflow ($\sqrt{-g} \rho U^r$) and effective potential
contours at the same late time in
simulation KDPg (left panel) and KDR (right panel)
as in Fig.~\ref{fig:angmomdist}.  Note the expanded
scale of this figure relative to that one.  Here, the color
contours are logarithm of the mass outflow rate, while
the line contours are linear in the potential
with separation 0.025.  The mean angular momentum used
to define the effective potential is 2.5 in the left panel and
4 in the right.  \label{fig:angmomequipot}}
\end{figure}

Because the mean specific angular momentum changes little through
the outflow, it makes sense to describe the outflow dynamics as taking
place with respect to a single effective potential
\begin{equation}
{\cal U} = \frac{\langle U_\phi \rangle_\rho}{g^{tt}}
           \left\{-g^{t\phi} + 
           \left[\left(g^{t\phi}\right)^2 - g^{\phi\phi}g^{tt}
           - g^{tt}/\left(\langle U_\phi \rangle_\rho\right)^2\right]\right\}.
\end{equation}
That is, ${\cal U}$ is the energy $U_t$ a test particle would have if
orbiting with $U_\phi = \langle U_\phi \rangle_\rho$ and $U_r =
U_\theta = 0$ at a given location.  We display the correspondence
between the mass outflow and the shape of this effective potential in
Figure~\ref{fig:angmomequipot}.  Two examples are presented, $a/M =
0.9$, for which $\langle U_\phi \rangle_\rho \simeq 2.5$, and $a/M =
-0.9$, for which $\langle U_\phi \rangle_\rho = 4$.   In both, the mass
flow must climb in the effective potential out to $r \simeq 10M$, but
``slides" at roughly constant potential from there outward.  Although
illustrated only for these two cases, this pattern is generally seen in
the other high-spin simulations as well.

\clearpage

\clearpage

It should not be too surprising that, for the most part, the outflow
moves through the region of least effective potential permitting
access to infinity.  Thus, the shape of the mass outflow (at least
in the inner few tens of gravitational radii) is determined by two
factors: its specific angular momentum (which is the angular momentum
of the inner disk) and the gravitational potential.  A corollary of the
importance of the effective potential is that there is very little mass
outflow (or mass) at polar angles closer to the axis than the lowest open
effective potential contour.  The angular momentum of the matter creates
a centrifugal barrier that prevents matter from filling in the funnel.
Since the only source of mass is the accretion disk, the funnel is empty.
In general, provided there is little matter of truly small angular
momentum anywhere in the system, we expect that the mass flux in {\it
all} such outflows should trace out a hollow cone, while the interior
of the cone should have very low density.

The sharp density gradient in polar angle at the inner edge of the
funnel wall outflow poses a difficulty for numerical simulations: an
extremely fine gridscale is required in the vicinity of the centrifugal
barrier.  The simulations reported here, while well-resolved in terms
of density elsewhere, fail to describe the sharpness of this gradient.
The fractional contrast in density from one cell to the next (moving
purely in $\theta$) on the inner edge of the funnel wall is typically
$\sim 1$--10.  Because of this poor resolution at the edge of the funnel,
the rest-mass and internal energy content of the funnel interior may not
be well-described quantitatively.  Because there is negligible matter
in the simulation with small enough angular momentum to enter the
funnel interior, anything found there came either from invocation of
the numerical density floor, or via numerical diffusion that resulted
in a low ratio of angular momentum to matter.  Of course this process
of angular momentum ``filtration'' should exist in real systems, since
only very low angular momentum material is allowed within the funnel.
Unfortunately, it is difficult to estimate how much truly low angular
momentum matter may be found close to the center of an accretion flow---so
little mass is needed in order to substantially change the content of
the funnel interior that many physical processes, justly neglected in
the rest of the simulation, may contribute.  For these reasons, we cannot
be quantitative regarding the matter content in the funnel interior.

\subsection{Mass outflow rate}

Although the mass density inside the funnel is ill-determined, the mass
density in the funnel wall, where substantially all of the unbound mass
outflow takes place, is much more reliable.  We can therefore examine
how the rate of mass loss in the outflow compares to the accretion
rate, and how this changes as a function of black hole spin.

However, there are certain cautions that attach even to this quantity.
First, the ratio of mass outflow rate to mass accretion rate is meaningful
only when averaged over long time intervals that exclude the initial
transients.   On the one hand, the accretion rate into the black hole is
highly variable, and the timescales of these variations are often quite
short, while on the other hand, the outflow varies more slowly because
its characteristic timescale $r_{\rm out}/c$ is comparatively long.
In addition, there are large amplitude transients in both the mass inflow
and outflow rates that typically occur $\sim 2000M$ after the beginning
of each simulation.  We therefore consider the ratio between the total
mass lost in in the outflow to the total rest-mass accreted over the
time 3000--$10^4 M$.

Second, although the density gradient within the funnel wall outflow is
modest, its angular width in terms of cells is only about half a
dozen.  Thus, there remains a noticeable numerical uncertainty in any
quantity integrated over the funnel wall.

Third, the distinction between unbound and bound mass flow is a subtle
one.  Wherever there is significant mass along the outer boundary, it
is moving outward.  The specific energy ($-hU_t$) varies smoothly with
polar angle at this boundary, so that while one can define a boundary
at $-hU_t = 1$, there is no sharp dividing line in energy between the
unbound and bound outflow near the funnel wall.

Finally, as shown in Paper~III, the mean unbound mass outflow rate
increases with increasing radius.  Beyond $r \simeq 20M$, the scaling
is slow (roughly $\propto r^{1/2}$), but it is clearly still varying
within our simulation volume.  Consequently, we stress that the numbers
given are the mass flows {\it at a fiducial large radius}, and {\it
not} necessarily the mass flow that would reach infinity.

The mass outflow data, subject as they are to all these {\it caveats},
are given in Table \ref{table:mflux}.  Very little mass is expelled
in the Schwarzschild black hole model.  When $a/M$ exceeds 0.9, the
rate begins to rise sharply.  It is noteworthy that the mass loss rate
relative to the accretion rate in the retrograde case, $a/M = -0.9$,
is almost the same as in the prograde case with the same magnitude
spin parameter.
There is little apparent trend between $a/M = 0.9$ and 0.95; for all these
simulations, the mass loss rate is $\simeq 8\%$  of the accretion rate.
The scatter between these numbers is likely an estimate of (or perhaps
a lower bound on) their uncertainty.  At the highest spin, 0.99, the
mass lost in the funnel wall outflow is nearly a quarter of the mass
accreted by the black hole.

Similarly, the average mass outflow velocity is at least partially a
function of black hole spin.  We compute an average velocity by the
ratio of the time-averaged funnel wall mass flux to the time-averaged
density in the funnel wall jet. As with many quantities, the value of
the velocity is a function of radius.  For the $a/M = 0$ model, the
mean funnel wall jet velocity is $v/c = 0.24$ at $r=100M$.  The
retrograde spin model has a comparable jet velocity, while the velocity
at $r=100M$ increases to $v/c = 0.3$--0.4 for the high spin models.

\subsection{Outward forces}

\begin{figure}
\psfig{file=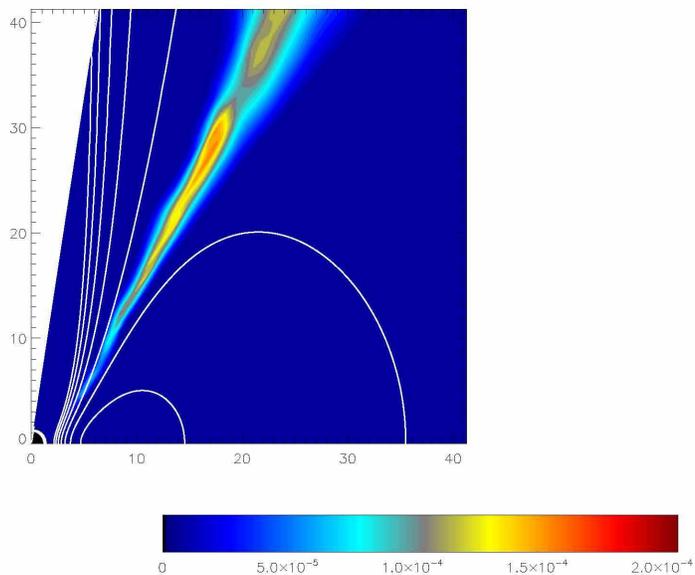,angle=90,width=5.5in}
\caption{Momentum outflow ${\cal P}$ and effective potential contours at
late time in simulation KDH ($a/M = 0.95$).  The color
contours are linear in ${\cal P}$, while
the line contours are linear in the potential for $U_\phi = 3.0$ (the
most appropriate $\langle U_\phi \rangle_\rho$ for this simulation)
with separation 0.05.\label{fig:momflux}}
\end{figure}

We now embark on a search, at a qualitative level, for the forces
responsible for driving this outflow.  The most elementary evidence in
this regard can be seen in Figure~\ref{fig:momflux}, which shows the
azimuthally-averaged matter momentum outflow in the funnel-wall, i.e.,
\begin{equation}
{\cal P} = (2/\pi)\int \, d\phi \, \sqrt{-g} \rho U^r U_r {\cal W}.
\end{equation}
We omit the factor $h$ from the expression for the momentum density
in order to suppress contributions from the funnel interior; $h\simeq
1$ everywhere in the funnel wall.  As this figure shows, the matter
momentum outflow in the funnel wall increases out to $r \simeq 20M$.
Time-dependent fluctuations cause the momentum outflow to vary with
increasing radius at larger distance, but without any consistent trend.
Moreover, the track of the momentum outflow follows a path along which
the effective potential hardly changes.  Figure~\ref{fig:angmomdist}
also provides evidence pointing toward the same conclusion.  The figure
illustrates that the mean specific angular momentum in the funnel-wall
hardly changes with radius.   Thus, the effect of any torques must be
quite minor.


\begin{figure}
\psfig{file=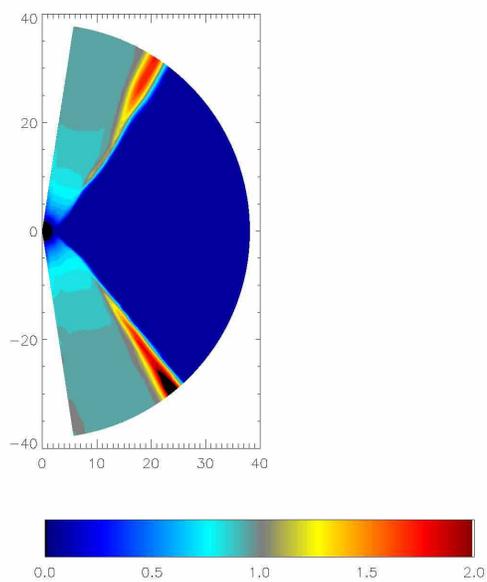,angle=90,width=5.5in}
\caption{Azimuthal average of the approximate magnetosonic Mach number
in matter belonging to the outflow
at a late time in the $a/M = 0.5$ simulation.\label{fig:mach}}
\end{figure}

These results indicate that while centrifugal effects help guide the
outflow, they do {\it not} accelerate it: other forces must be responsible
for its radial acceleration.  To ascertain just what these forces are,
we look at the problem from a different point of view.   In Paper~III,
we demonstrated that mass in the outflow is acquired rapidly in the
range between a few $r_{ms}$ and $\simeq 20M$, but continues to grow
slowly with increasing radius beyond that point.   In order for mass
to be introduced to the outflow over such a broad range of radii,
the acceleration region must be distributed along the length of the
funnel wall.  Just such a situation can be seen in Figure~\ref{fig:mach}.
In that figure, we show a Newtonian approximation to the Mach number
relative to the magnetosonic speed, i.e., $V^r/c_{\rm ms}$, where
$c_{\rm ms}^2 = [||b||^2 + (5/3)p]/(||b||^2+h\rho)$.  Regions outside
the unbound outflow are suppressed in this figure.   From $r \simeq 10M$
to the outside, the transonic contour follows the outside edge of the
funnel-wall.  That is, fresh matter is both injected and accelerated
along the length of the jet.  In much of the funnel interior, the flow
is sub-magnetosonic because $||b||^2/\rho$ is so large that $c_{\rm ms}
\simeq 1$; at larger radii in the funnel interior, $V^r \rightarrow 1$,
so that, in this sense, the flow approaches Mach number unity.

\clearpage

\clearpage

\begin{figure}
\psfig{file=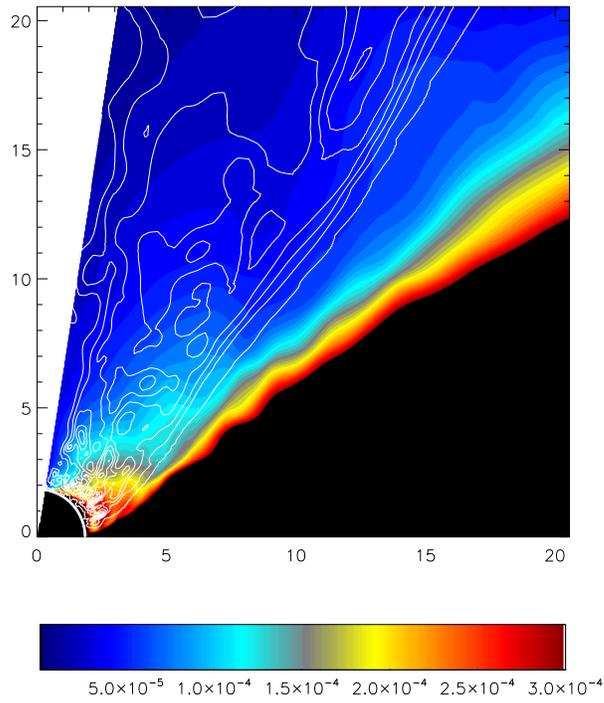,angle=90,width=5.5in}
\caption{Azimuthal average of the total pressure (gas plus
magnetic) in color contours at the same time in the $a/M=0.5$
simulation shown in Fig.~\ref{fig:mach}.  Radial momentum
flux in the outflow is shown in white contours.  Both sets
of contours are linear
scales.  The highest contour in momentum flux is roughly
twice the highest pressure shown in the color scale.
\label{fig:pressuredist}}
\end{figure}

Pressure forces account for this distributed acceleration.
Figure~\ref{fig:pressuredist} shows the total pressure
(more precisely, $\sqrt{-g}(||b||^2/2 + p)$) compared to
the radial matter momentum flux in the outflow ($\sqrt{-g}\rho h U^r U_r$).
The pressure contours are oblique to the momentum flux
contours, angled at such a way as to push matter into the
outflow as well as to accelerate it radially.  Matter is prevented
from moving deeper into the funnel interior by the sharp rise
in the effective potential (cf. Fig.~\ref{fig:momflux}).  In the
coronal region generally, and near the funnel wall in particular,
the magnetic and gas pressures are comparable to one another.
Interestingly, the contours of both the gas pressure and the magnetic
pressure considered alone are significantly more irregular
than the contours of their sum.


\subsection{Collimation}

We have already demonstrated that the inner edge of the matter
outflow is determined by the centrifugal barrier in the inner few
tens of gravitational radii, and that its flow beyond that distance
is essentially ballistic within the nearly-flat effective potential.
The electromagnetic fields in the funnel interior expand to fill the
hollow region, so the matter's dynamics determine the edge of the
electromagnetically-dominated funnel interior.  The outer edge of the
matter outflow (funnel wall jet) is less sharply marked.  The specific
energy ($-h U_t$) is a smooth function of polar angle at large radius, and
there is no sharp division between bound and unbound regions.  Similarly,
the pressure is likewise smoothly varying (cf. Paper~III).  In this
sense, we consider the outflow to be collimated by the coronal pressure.
For this reason, future work with more realistic coronal thermodynamics
could reveal interesting changes in outflow shape.
However, the smoothness of the border between outflow and bound corona
also means that it is ill-marked.  Indeed, the rise of the mass outflow
rate with radius may be an indication that entrainment plays a significant
role in mass entry at large radius, further blurring this edge.

\begin{figure}
\psfig{file=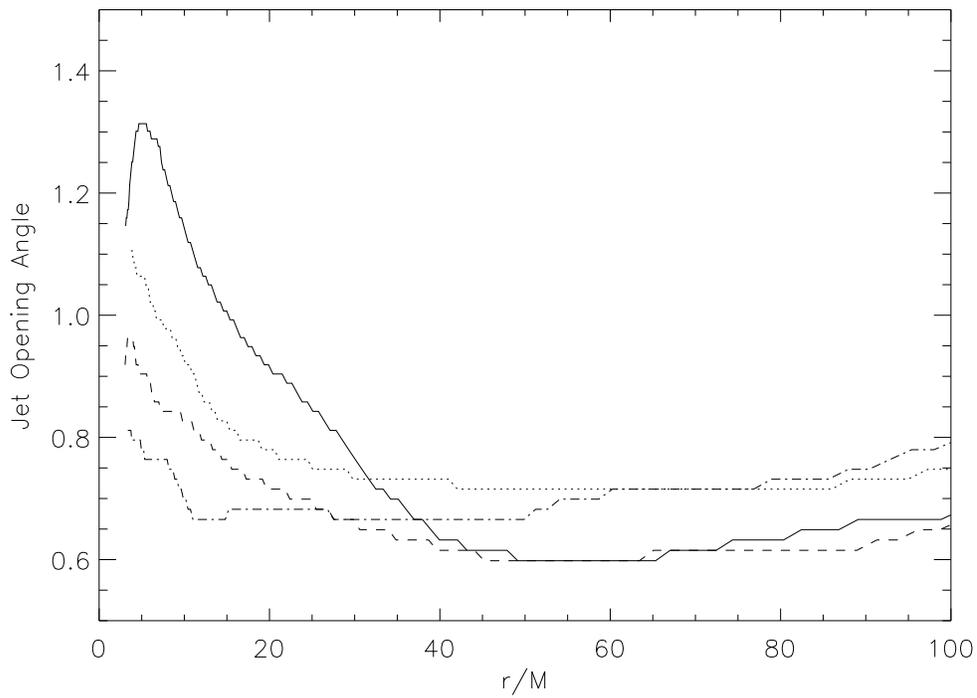,angle=90,width=5.5in}
\caption{
Opening angle of the jet in radians, as defined by the angular location of
the boundary between bound and unbound outflow.  Four models are shown
at $t=10,000M$:
the solid line is $a/M = -0.9$, the dotted line is $a/M = 0.5$, 
the dashed line is $a/M = 0.9$, and the dot-dashed line is $a/M = 0.99$. 
\label{fig:collimation}}
\end{figure}

To quantify the collimation, we plot in Figure~\ref{fig:collimation}
the opening angle of the jet obtained by locating the angular
boundary between the bound and unbound flow as a function of radius.
As the figure shows, the higher the spin, the more narrow the jet
initially, although all of the jets collimate out to around
$r=40$--$50M$.  Beyond that point, the opening angle remains constant
or increases very slightly toward the outer boundary of the simulation
domain.  The spin-dependence of the opening angle at small radii is
most likely due to the fact that the stable disk extends farther inward
with increasing spin.  Where there is stable disk in the equatorial
plane, it is easier to support a high-pressure corona.


\section{Summary and Discussion}

In this paper we have examined the unbound axial outflows (jets)
that develop within GRMHD accretion simulations.  What is noteworthy
in these models is that the jets develop as a natural result of the
accretion process, rather than as a direct consequence of conditions
imposed on the simulation.  The jets have two major components:  a
matter-dominated outflow that moves at a modest velocity ($v/c \sim 0.3$)
along the centrifugal barrier surrounding an evacuated axial funnel, and
a highly-relativistic Poynting flux dominated jet within the funnel.
The funnel wall jet is accelerated and collimated by magnetic and
gas pressure forces in the inner torus and the surrounding corona.
The Poynting flux jet results from the formation of a large scale radial
magnetic field within the funnel.  This field is spun by the rotating
spacetime of the black hole, hence the energy for this jet component comes
not directly from accretion but from the black hole's rotation, although
the field mediating this energy extraction is a product of accretion.
The energies in both components of the jet are a function of black hole
spin; greater spins yield greater energy.  The total jet power can be
a significant fraction of the accretion energy estimated by the product
of the mass accretion rate and the specific binding energy of the last
stable orbit.

We next consider the broader implications of these results
within the context of several of the standard theoretical categories
for jets, namely jet acceleration, power, collimation, and the
Blandford-Znajek mechanism.

\subsection{Accelerating forces}

It has been expected for many years (since Blandford \& Znajek 1977),
that within the axial funnel the spacetime rotation enforced by spinning
black holes can drive a significant electromagnetic outflow.  Our
results are clearly in line with that expectation.

The outflow-generation process within the accretion flow itself that has
for many years received the greatest attention has been magnetocentrifugal
acceleration, such as envisioned in the model of Blandford \& Payne
(1981), but this mechanism does not appear to be important here.

On the other hand, the main driving force we identify for the funnel
wall jet---a high-pressure corona squeezing material against an inner
centrifugal wall---has received very little attention in the past.
In consequence of this compression, the pressure at the base of the
matter outflow is kept high enough to create an outward radial pressure
gradient that provides the momentum for the outflow.  It is also worth
commenting that these forces act over a substantial radial range, from
a few gravitational radii out to many tens.

\subsection{Outflow strength}

To measure the relative power in the unbound outflow, we have compared
the jet energy flux with the mass accretion rate into the hole over the
last 70\% of the simulation.  The simulation has no radiative losses,
so we cannot say what the radiative efficiency of the accretion disk
would be.  The Novikov-Thorne efficiency provides a convenient standard
of comparison.  This nominal efficiency corresponds to the binding
energy of the innermost stable circular orbit, although the actual
radiative efficiency could well be different because of, for example,
magnetic torques in the plunging region (Krolik 1999; Gammie 1999), or
heat advection into the hole.  Judged by this measure, the energy
carried off in the outflow is considerable.  In two cases ($a/M = -0.9$
and 0.99), the electromagnetic power at $r=100M$ is, by itself,
comparable to the Novikov-Thorne prediction of radiated power.  In the
other high spin cases, it is about a quarter of the Novikov-Thorne
number.  One should remember, however, that energy can be (and often
is) exchanged between the fields and the matter, so these measures of
the electromagnetic power in the outflow are quantitatively accurate
{\it only} at the fiducial location where they are measured.  Because
the rate of energy loss in the matter outflow is usually a few times as
great as the electromagnetic contribution, the total is typically
comparable to the classical prediction of radiated power.

Another way to gain a feel for the magnitude of the electromagnetic power
in the outflow is via dimensional analysis.  The Poynting flux at the
inner boundary is
$\propto {\cal B}^r {\cal B}^\phi r_{\rm in} \omega_{\rm in}$
(cf. eqn.~\ref{eqn:bzpoynting}); but what determines the magnetic field
strength?  Indeed, this has been the key question examined by such works
as Ghosh \& Abramowicz (1997) and Livio, Ogilvie, \& Pringle (1999).
These authors argued that the strength of the field responsible for a
Blandford-Znajek process must be set by the fields in the accretion
flow, and this may limit the total jet power.

\begin{figure}
\centerline{\psfig{file=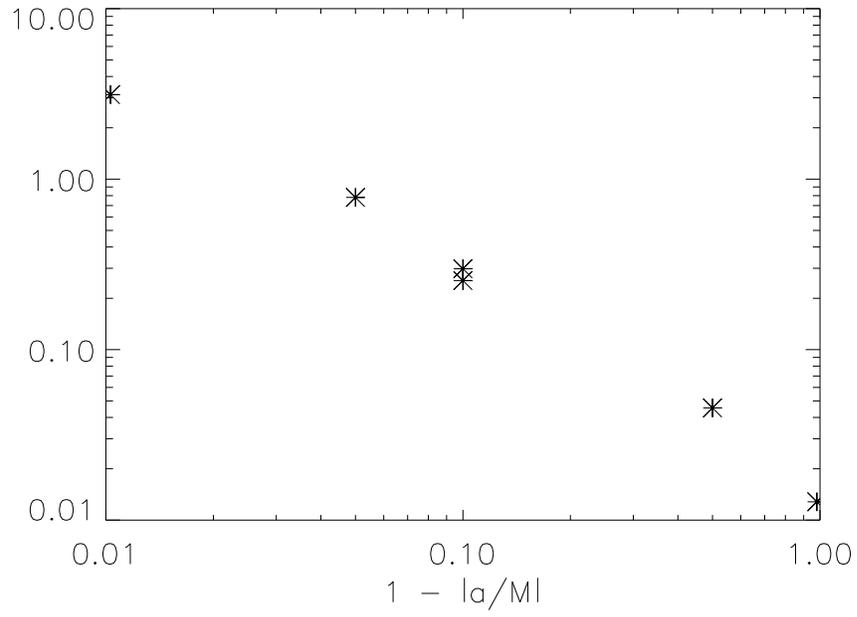,angle=90,width=5.0in}}
\caption{The ratio $||b||^2/\dot M$ averaged over time, with
the magnetic field intensity evaluated at the inner boundary.
\label{fig:bsqeff}}
\end{figure}

At radii in the disk well beyond the marginally stable region, angular
momentum conservation ensures that
$\langle {\cal B}^r {\cal B}^\phi \rangle \propto \dot M \Omega/h$ if
magnetic torques are the primary agent of accretion, while the
competition between orbital shear and other stresses tends to make
$\langle ||b||^2 \rangle \propto \langle {\cal B}^r {\cal B}^\phi 
\rangle$.  In these
simulations, in which $h/r$ is common to all, but the black hole spin
varies, if we wish to analyze the spin-dependence of the Poynting flux
near the event horizon, it makes sense to examine the ratio of the
magnetic field intensity to the mass accretion rate at that location.
We find that the time- and surface area-averaged ratio $||b||^2/\dot M$
grows rapidly with increasing black hole spin, very nearly $\propto (1
- |a/M|)^{-1.1}$ (Fig.~\ref{fig:bsqeff}).  Some of this growth is
attributable to mere compression, as the event horizon shrinks as the
spin rate rises.  However, the dependence is much stronger than
anything simple geometry could explain.  It appears that the strong
frame-dragging associated with extreme Kerr spacetimes leads to
substantial field amplification as well.  The rapid growth in
electromagnetic power output with faster black hole rotation is
predominantly due to this effect.

\clearpage

\clearpage

\subsection{Jet shape}

One of the key issues in jet formation and propagation is the degree
to which the jets are collimated.  Of course, jets are observed
on far larger scales than we are modeling in these simulations, but
the degree of collimation seen here is nevertheless of some interest.
Generally speaking, whatever collimation there is occurs in the inner
regions of the flow, where the inner disk and hot corona provide
confinement.  Relatively close to the black hole ($r \lesssim 10M$),
the boundary between the magnetically-dominated Poynting flux jet and
the funnel wall outflow follows the shape of the centrifugal barrier.
At larger distances, however, the matter flows almost straight
outward, so that the funnel is nearly conical, while the contours of
constant effective potential gradually move toward smaller polar angle
(Fig.~\ref{fig:momflux}).

The precise shape and collimation of the jet is somewhat uncertain
because the outer boundary of the funnel-wall, matter-dominated funnel
wall outflow is somewhat indistinct.  We see a smooth transition as a
function of polar angle between mildly relativistic outflowing unbound
matter and slightly slower, but bound, coronal matter.  On the other
hand, the boundary between the low-density magnetically-dominated funnel
interior and the higher-density funnel-wall outflow is sharp and clear.
It is possible that magnetic forces might provide additional collimation
on larger scales, but this remains to be seen in future simulations.

\subsection{Comparison to other models and the ``Blandford-Znajek" mechanism}

In the astrophysical literature, the general idea that energy can
be conveyed from rotating black holes to the outside world via
magnetic connections is often dubbed the ``Blandford-Znajek mechanism"
because they were the first to discuss a specific version of this
process.  In our simulations, when all is said and done, the funnel
interior is dominated by a magnetic field that would be purely radial
but for the rotation created by its connection to the black hole's
ergosphere.   Thus, any toroidal field in the funnel, and consequently
any Poynting flux, is due entirely to general relativistic dynamics
peculiar to spinning black holes.  In other words, the outward
Poynting flux must be due to a process that could fit under the
general rubric of ``Blandford-Znajek".

On the other hand, we also find specific differences from the classical
Blandford-Znajek model, specifically the particular idealized picture
presented in their original paper (Blandford \& Znajek 1977).  For
example, although the funnel region is magnetically-dominated, it is
{\it not} in general in a state of force-free equilibrium.  Indeed, the
very large fluctuations that continually occur in the outflow show that
it is never in any state of equilibrium, force-free or otherwise.  In
addition, as originally shown in Phinney (1983), the collapse of all
wave speeds to $c$ in the exact force-free limit means that to
understand the dynamics properly requires retention of a finite
inertia.

Moreover, whereas in the classical Blandford-Znajek model there was
zero accretion (the only function of nearby orbiting matter is to
support the currents sustaining the magnetic field), in our picture the
electromagnetic luminosity in the outflow takes place entirely as a
byproduct of accretion.  In the classical model the magnetic field was
a relic of plasma inflow that took place in the distant past; because
there was no continuing accretion, the energy output could be
attributed unambiguously to the black hole's rotation.  In our
simulation, the magnetic field is a direct result of on-going
accretion.  While the proximate energy source for the jet is the black
hole's rotation, accretion replenishes both the black hole's mass and
angular momentum, and whether the rotational energy of the black hole
actually decreases depends on details such as the radiative efficiency
and, possibly, the accretion of photons (Krolik 2001; cf. Thorne
1974).  In our $a/M = 0.99$ simulation, the net angular momentum
accreted is so small that, for reasonable values of the radiative
efficiency, the system is very near the edge of an actual decrease in
rotational kinetic energy.  Even if the black hole's rotational energy
does not decrease, there can nevertheless be a significant increase in
the efficiency of accretion.  We find that for rapidly rotating black
holes the total power in the Poynting flux jet can be comparable to the
nominal accretion power.

A number of other numerical studies of the generic Blandford-Znajek
mechanism have been carried out recently.  One type of simulation has
been devoted to examining the idealized problem of a rotating black
hole embedded in thin plasma containing a specified external magnetic
field (Koide 2003; Semenov et al. 2004; Komissarov 2005).  These
simulations focus on points of basic principle, such as the extent to
which accretion of either matter or electromagnetic fields with
negative energy-at-infinity plays a role.  Komissarov (2005)
particularly notes that plasma effects within the ergosphere were
significant early on in his simulation while the initial plasma is
accreting into the hole, but that the long-term steady state mechanism
resembles closely the classic Blandford-Znajek model.

The simulations of McKinney \& Gammie (2004) were more similar in their
focus to ours, as they too were principally concerned with the
connection between continuing accretion and the electromagnetic power
output of black holes.  For this reason, their (axi-symmetric)
simulations employed initial conditions resembling ours (isolated
initial gas torus containing field loops).  They found good agreement
with several aspects of the classical Blandford-Znajek model for slowly
rotating holes, but growing differences as the spin of the hole
increases and the field becomes less monopolar.  That this should be so
is not too surprising, as the specific predictions of the
Blandford-Znajek model were developed on the basis of a perturbation
expansion in $a/M$.  In regard to the global energy and angular
momentum budgets, McKinney \& Gammmie's results qualitatively resemble
ours.  They also saw that as the hole loses energy and angular momentum
electromagnetically, its total mass and angular momentum are
replenished by accretion.  Likewise, they, too, found that the overall
efficiency (energy released to infinity per rest-mass accreted) can be
significantly enhanced by the energy carried off in the jet.  McKinney
(2005) expanded upon these results and proposed an analytic fitting
function for the electromagnetic efficiency of the jet as a function of
black hole rotation rate, suggesting that this efficiency scales
$\propto \Omega_H^5$, where $\Omega_H$ is the rotation rate of the
black hole.  If the Poynting flux were due simply to rotation of a
magnetic dipole whose magnitude is independent of black hole
rotation but $\propto \dot M^{1/2}$, the efficiency would, of course,
increase $\propto \Omega^4$.  The slightly faster dependence McKinney (2005) 
found suggests either that there is a mixture of a higher multipole
component or that the magnitude of the magnetic dipole increases slowly with
increasing black hole spin.

The choice of this functional form was motivated by
the hope that electromagnetic radiation in this case resembled
classical dipole radiation, in which the luminosity
is $\propto B^2 \Omega^4$.   Whether this applies in the present context
depends, of course, on whether dipole field geometry is a reasonable
approximation and on whether the mean field strength
at the horizon produced by
a fixed accretion rate is independent of  spin.

Unfortunately, for a variety of technical reasons it is difficult to
compare quantitatively the results of different numerical simulations.
Without the inclusion of additional physics in the basic equations,
there is no absolute density scale in the gas.  Although efficiencies
are well-defined (ratios of energies to masses within the simulation),
one cannot define an absolute accretion rate.  Even the comparison of
efficiencies from one type of simulation to another poses problems.
For example, two dimensional simulations tend to over-emphasize
axisymmetric MRI modes.  As a result, they tend to have large accretion
rates early on, followed by a rapid decline as the magnetic field and
accompanying turbulence die out, as expected from the anti-dynamo
theorem.  Thus, when computing the efficiency in a two dimensional
simulation simulation, one must be careful about choosing an
appropriate interval over which to average the accretion rate.  A
second problem is that, as discussed previously, energy can
be exchanged between the electromagnetic and matter components of the
outflow, so the purely electromagnetic power is a function of radius.
To cross-compare different simulations, it is therefore important to
use the same fiducial distance for evaluating the electromagnetic
power.   Lastly, jet properties may depend on the vertical thickness of
the accretion disk, a parameter determined by the initial temperature
of the disk and the treatment of the energy equation in the
simulation.  The gross accretion rate in general scales $\propto h^2$,
but we do not know how the magnetic field in the vicinity of the black
hole scales with disk temperature.  Further simulational work will be
required to determine such effects, although because the strength of
the axial funnel field is determined by the accretion flow, it is a
reasonable expectation that the jet power might scale with the
accretion rate (Ghosh \& Abramowicz 1997).

Keeping these {\it caveats} in mind, we can compare our electromagnetic
efficiency in the jet with the
scaling relation proposed by McKinney (2005) by modifying his relation
to use the rotation rate of our inner grid radius instead of the
rotation rate of the horizon.  The quantitative match between our
data and McKinney's formula is not particularly good, although
qualitatively we, too, see a strong increase in efficiency as black hole spin
increases.  Our results are better described by the simple formula
\begin{equation} 
\eta_{em} \simeq 0.002/(1 - |a/M|) , 
\end{equation} 
a form that describes both the high and lower spin models.  Of
course, without any clear theory to predict the strength of the magnetic
field on the event horizon relative to the mass accretion rate as a
function of black hole rotation rate, one cannot expect to
find a general formula for total jet power.

Finally, we would like to close this paper by pointing out that,
although the Blandford-Znajek mechanism was originally thought of as a
scheme for powering jets, outflows are not the only place where
electromagnetic power from the black hole can go.  In this paper we
have concentrated on jet properties.  However, preliminary analysis of
the accretion flow in these simulations indicates that the Poynting
flux and electromagnetic angular momentum flux from the black hole into
the disk can be comparable to those within the outflow.  These fluxes
and their consequences will be examined in subsequent work.

\acknowledgements{
This work was supported by NSF grant
PHY-0205155 and NASA grant NNG04GK77G (JFH),
and by NSF grants AST-0205806 and AST-0313031
(JHK).  We acknowledge Jean-Pierre De Villiers for continuing 
collaboration in the
development of the algorithms used in the GRMHD code.  
The simulations were carried out on the DataStar
system at SDSC.  A portion of this work was carried out while
the authors were at the Kavli Institute for Theoretical Physics,
supported by NSF grant PHY99-07949.
}


\clearpage

\begin{deluxetable}{lrrrrrrr}
\tablecolumns{6}
\tablewidth{0pc}
\tablecaption{\label{table:sims} Simulation Parameters}
\tablehead{\colhead{Name}          &
           \colhead{$a/M$}         &
           \colhead{$r_{+}/M$} &
           \colhead{$r_{\rm in}/M$} &
           \colhead{$W_{\rm max}$} &
           \colhead{$\omega(r_{\rm in})$} &
           \colhead{$r_{\rm ms}/M$}  &
           \colhead{$U_\phi(r_{\rm ms})/M$}
}
\startdata
KDR  & -0.9  & 1.44 & 1.503  & 10. & -0.287 & 8.72 & 4.17  \\
KD0c & 0.0   & 2.00 & 2.104  & 10. & 0.0      &6.00 & 3.46  \\
KDIb & 0.5   & 1.86 & 1.904  & 10. & 0.127   &4.23 & 2.90  \\
KDPg & 0.9   & 1.44 & 1.503  & 10. & 0.289 &2.32 & 2.10  \\
KDG  & 0.93  & 1.37 & 1.458  & 6.  & 0.306  &2.10 & 1.98  \\
KDH  & 0.95  & 1.31 & 1.403  & 6.  & 0.326  &1.94 & 1.89  \\
KDJa & 0.99  & 1.14 & 1.203  & 6.  & 0.406 &1.45 & 1.57  \\
\enddata
\tablecomments{Here $a/M$ is the spin parameter of the black hole,
$r_{+}$ is the horizon radius,
$r_{\rm in}$ is the innermost radius in the computational grid,
$W_{\rm max}$ is the largest Lorentz factor permitted, $\omega(r)$
is the frame-dragging rate at that radius in the equatorial plane
(i.e., the rotation rate of a zero angular-momentum observer),
$r_{\rm ms}$
is the radial coordinate of the innermost stable circular orbit
in the equatorial plane, and $U_\phi(r_{\rm ms})$ is the specific angular
momentum of that last stable orbit.}
\end{deluxetable}

\begin{deluxetable}{lrrrrr}
\tablecolumns{6}
\tablewidth{0pc}
\tablecaption{\label{table:mflux} Jet Mass Flux }
\tablehead{\colhead{Model} &
\colhead{$a/M$} &
\colhead{$M_{\rm tot}$} &
\colhead{${M_{\rm hole}}$} &
\colhead{${M_{\rm jet}}$} &
\colhead{${\dot M_{\rm jet}/\dot M_{\rm hole} }$} } 
\startdata
KDR  &-0.90 & 195.6 & 22.1 & 0.77 & 0.035 \\
KD0c & 0.00 & 156.3 & 22.2 & 0.11 & 0.005 \\
KDIb & 0.50 & 234.8 & 20.6 & 0.76 & 0.037 \\
KDPg & 0.90 & 290.5 & 19.2 & 1.71 & 0.089 \\
KDG  & 0.93 & 319.0 & 21.8 & 1.72 & 0.079 \\
KDH  & 0.95 & 339.2 & 20.3 & 1.72 & 0.085 \\
KDJ  & 0.99 & 382.5 & 15.5 & 3.66 & 0.237 \\
\enddata
\tablecomments{$M_{\rm tot}$ is the total mass in the
initial hydrostatic torus; $M_{\rm hole}$ is the total
rest-mass passing through the inner boundary in
the time interval 3000--10000$M$; $M_{\rm jet}$ is
the total unbound rest-mass leaving the simulation
region over the same interval.  All masses are given
in the (arbitrary) code mass-units.}
\end{deluxetable}

\begin{deluxetable}{lrrrrr}
\tablecolumns{6}
\tablewidth{0pc}
\tablecaption{\label{table:lflux} Angular Momentum Flux }
\tablehead{\colhead{Model} &
\colhead{$a/M$} &
\colhead{${{L^{m}}_{\rm hole}}$} &
\colhead{${{L^{em}}_{\rm hole}}$} &
\colhead{${{L^{m}}_{\rm jet}}$} &
\colhead{${{L^{em}}_{\rm jet}}$} }
\startdata
KDR  &-0.90 & -82.7 & -4.8  & 3.15 & -2.65 \\
KD0c & 0.00 & -72.4 &  1.4  & 0.31 & 0.14 \\
KDIb & 0.50 & -57.0 &  2.6  & 1.40 & 2.10 \\
KDPg & 0.90 & -42.9 &  7.5  & 2.26 & 6.96 \\
KDG  & 0.93 & -46.0 &  11.7 & 5.44 & 6.36 \\
KDH  & 0.95 & -42.2 &  16.8 & 5.68 & 8.88 \\
KDJ  & 0.99 & -28.7 &  26.4 & 14.3 & 12.8 \\
\enddata
\tablecomments{
$L^{m}_{\rm hole}$ is the total
angular momentum carried by the mass passing through the inner boundary in
the time interval 3000--10000$M$; $L^{em}_{\rm hole}$ is the angular momentum 
carried by the electromagnetic field.  The values marked ``jet''
are the total angular momenta leaving the simulation
region over the same interval in the unbound jet outflow. 
}
\end{deluxetable}

\begin{deluxetable}{lrrrrrrrr}
\tablecolumns{6}
\tablewidth{0pc}
\tablecaption{\label{table:eflux}
  Energy Flux }
\tablehead{\colhead{Model} &
\colhead{$a/M$} &
\colhead{$E^{m}_{\rm hole}$} &
\colhead{$E^{em}_{\rm hole}$} &
\colhead{$E^{m}_{\rm jet}$} &
\colhead{$\eta_{m}$}&
\colhead{$E^{em}_{\rm jet}$}&
\colhead{$\eta_{em}$}&
\colhead{$\eta_{\rm NT}$}  }
\startdata
KDR  &-0.90 & -20.8 &  0.41 & 2.71 & 0.088 & 0.509 & 0.023 & 0.039\\
KD0c & 0.00 & -20.6 & -0.02 & 0.16 & 0.0022 & 0.007 & $3.1 \times10^{-4}$&0.057 \\
KDIb & 0.50 & -21.7 &  0.10 & 2.06 & 0.063 & 0.129 & 0.0063 & 0.081\\
KDPg & 0.90 & -21.8 &  0.88 & 5.89 & 0.22  & 0.892 & 0.046  & 0.155\\
KDG  & 0.93 & -19.8 &  1.37 & 3.13 & 0.065 & 0.824 & 0.038  & 0.173 \\
KDH  & 0.95 & -16.9 &  2.79 & 4.26 & 0.13  & 1.46  & 0.072  & 0.190\\
KDJ  & 0.99 &  -9.9 &  6.86 & 9.94 & 0.41  & 3.28  & 0.21   & 0.264 \\
\enddata
\tablecomments{Negative signs for $E_{\rm hole}$ signify energy
flows into the black hole; positive signs for that quantity
indicate energy flowing into the computational domain.
For definitions of the several efficiencies denoted
by $\eta$ with different subscripts, see the text.}
\end{deluxetable}

\begin{thebibliography}

\bibitem[Balbus \& Hawley (1998)]{bh98} Balbus, S.~A. \& Hawley, J.~F.
 1998, Rev. Mod. Phys., 70, 1 
\bibitem[Blandford \& Znajek (1977)]{bz77} Blandford, R.~D. \&
 Znajek, R. 1977, MNRAS, 179, 433 
\bibitem[Blandford \& Payne (1982)]{bp82} Blandford, R.~D. \& Payne,
 D. 1981, MNRAS, 199, 833 
\bibitem[De Villiers, \& Hawley (2003)]{DH03a} De Villiers, J.-P. \& 
 Hawley, J.~F. 2003, ApJ, 589, 458  
\bibitem[De Villiers, Hawley, \& Krolik (2003)]{PaperI} De Villiers, J.-P., 
 Hawley, J.~F., \& Krolik, J.~H. 2003, ApJ, 599, 1238 (Paper~I) 
\bibitem[De Villiers, Hawley, Krolik \& Hirose (2005)]{PaperIII} De Villiers, J.-P., 
 Hawley, J.~F., Krolik, J.~H. \& Hirose, S. 2005, ApJ, 620, 878 (Paper III)
\bibitem[De Villiers, Staff \& Ouyed (2005)]{dso05} De Villiers,
J.-P., Staff, J., \& Ouyed, R., astro-ph 0502225 
\bibitem[Evans \& Hawley]{eh88} Evans, C.~R., \& Hawley, J.~F., ApJ, 332, 659 
\bibitem[Gammie (1999)]{g99} Gammie, C.~F. 1999, ApJ Lett 522, L57 
\bibitem[Gammie et al. (2003)]{G03} Gammie, C.~F., McKinney,
  J.~C., \& T\'oth, G. 2003, ApJ, 589, 444 
\bibitem[Ghosh \& Abramowicz (1997)]{ga97}Ghosh, P. \& Abramowicz, M.~A. 1997,
MNRAS, 292, 887
\bibitem[Hirose, et al. (2004)]{PaperII} 
 Hirose, S., Krolik, J.~H., De Villiers, J.~P., \& Hawley, J.~F. 2004, ApJ, 
 606, 1083 (Paper~II) 
\bibitem[Kato, et al. (2004)]{K04} Kato, Y., Mineshige, S., \&
 Shibata, K. 2004, ApJ, 605, 307 
\bibitem[Koide (2003)]{koi03} Koide, S. 2003, Phys Rev D, 67, 104010
\bibitem[Komissarov (2004)]{ko04}Komissarov, S.~S. 2004, MNRAS, 350, 1431 
\bibitem[Komissarov (2005)]{ko05}Komissarov, S.~S. 2005, MNRAS, 359, 801
\bibitem[Krolik (1999)]{k99} Krolik, J.~H. 1999, ApJ Lett 515, L73 
\bibitem[Krolik (2001)]{k01} Krolik, J.~H. 2001, in {\it Explosive
Phenomena in Astrophysical Compact Objects}, I. Yi and M. Rho,
eds. (AIP: New York), p. 10
\bibitem[Krolik et al. (2005)]{PaperIV} Krolik, J.H., Hawley, J.F. \&
Hirose, S. 2005, ApJ 622, 1008 (Paper IV)
\bibitem[Livio, Ogilvie \& Pringle (1999)]{Liv99} Livio, M., Ogilvie, G.~I., \& Pringle, J.~E. 1999, ApJ, 512, 100 
\bibitem[Lynden-Bell (2003)]{LB03} Lynden-Bell, D. 2003, MNRAS 341, 1360 
\bibitem[McKinney (2005)]{mc05}McKinney, J.~C. 2005, ApJ, 630, L5
\bibitem[McKinney \& Gammie (2004)]{mg04} McKinney, J.~C., \& Gammie, C.~F.
 2004, ApJ 611, 977 
\bibitem[Novikov \& Thorne (1973)]{NT73} Novikov, I.D. \& Thorne, K.S.  1973,
 in Black Holes: Les Astres Occlus, eds. C. de Witt \& B. de Witt (New York:
 Gordon \& Breach), 344
\bibitem[Phinney (1983)]{P83} Phinney, E.S. 2003, unpublished Cambridge
University Ph.D. thesis 
\bibitem[Punsly \& Coroniti (1990)]{pc90} Punsly, B., \& Coroniti, F.~V.
 1990, ApJ, 354, 583 
\bibitem[Semenov, Dyadechkin \& Punsly (2004)]{sdp04} Semenov, V.,
Dyadechkin, S., \& Punsly, B. 2004, Science, 305, 978
\bibitem[Shibata \& Uchida (1985)]{su85} Shibata, K. \& Uchida, Y.
 1985, PASJ, 37, 31 
\bibitem[Stone \& Norman (1992a)]{sn2a} Stone, J.~M., \& Norman,
M.~L. 1992a, ApJS, 80, 753 
\bibitem[Stone \& Norman (1992b)]{sn92b} Stone, J.~M., \& Norman,
M.~L. 1992b, ApJS, 80, 791 
\bibitem[Thorne (1974)]{thorne74} Thorne, K.~S. 1974, ApJ, 191, 507


\end{thebibliography}
\end{document}